\documentclass[aps,superscriptaddress,nofootinbib,showpacs,eqsecnum,preprint,tightenlines]{revtex4}
\usepackage{hyperref}
\usepackage{epsfig,rotating}
\usepackage{amsmath,amssymb}
\usepackage{dsfont}
\numberwithin{equation}{section}

\newcommand{\vx}{\vec{x}}

\newcommand{\vp}{\vec{p}}
\newcommand{\vq}{\vec{q}}
\newcommand{\vk}{\vec{k}}

\newcommand{\oO}{E_\nu}

\newcommand{\be}{\begin{equation}}
\newcommand{\ee}{\end{equation}}
\newcommand{\bea}{\begin{eqnarray}}
\newcommand{\eea}{\end{eqnarray}}

\begin{document}

\title{Short baseline neutrino oscillations: when entanglement suppresses coherence.}

\author{Daniel Boyanovsky}
\email{boyan@pitt.edu} \affiliation{Department of Physics and
Astronomy, University of Pittsburgh, Pittsburgh, PA 15260}

\date{\today}

\begin{abstract}
For neutrino oscillations to take place the entangled quantum state of a neutrino and a charged lepton produced via charged current interactions must be disentangled.   Implementing a non-perturbative Wigner-Weisskopf method we obtain the correct \emph{entangled} quantum state of neutrinos and charged leptons from the (two-body) decay of a parent particle.  The source lifetime  and disentanglement length scale  lead to a suppression of the oscillation probabilities in  short-baseline  experiments. The suppression   is determined by    $\pi\, L_s/L_{osc}$ where $L_s$ is the \emph{smallest} of the decay length of the parent particle or the disentanglement length scale. For $L_s \geq L_{osc}$   coherence and oscillations   are suppressed. These effects are more prominent in \emph{short base line experiments} and at low neutrino energy.     We obtain the corrections to the   appearance and disappearance probabilities modified by both the lifetime of the source and the disentanglement scale and discuss their implications for accelerator and reactor experiments.  These effects imply that fits to the experimental data based on the usual quantum mechanical formulation \emph{underestimate} $\sin^2(2\theta)$ and $\delta m^2$, and are more dramatic for $\delta m^2\simeq \,\mathrm{eV}^2$,   the mass range   for new generations of sterile neutrinos that could explain the short-baseline anomalies and long disentanglement length scales.

\end{abstract}


\maketitle

\section{Introduction}

Neutrino masses, mixing and oscillations are the clearest evidence yet of physics beyond the standard model \cite{book1,book2,book3}. They provide an explanation of the solar neutrino problem \cite{msw,book4,haxtonsolar} and have   important phenomenological \cite{book1,book2,book3,grimuslec,kayserlec,mohapatra,degouvea,bilenky}, astrophysical \cite{book4,book5,haxton} and cosmological \cite{dolgovcosmo} consequences. A remarkable series of experiments have confirmed  mixing and  oscillations among three ``active'' neutrinos with $\delta m^2 = 10^{-4}-10^{-3}\,\mathrm{eV}^2$ for atmospheric and solar oscillations respectively\cite{revius}.

A fascinating aspect of neutrino oscillations is that they provide an extraordinary example of macroscopic quantum coherence maintained over hundreds of kilometers. It is particularly this aspect that has sparked an ongoing discussion in the field that seeks to clarify the main concepts behind the physical interpretation of oscillations.

As neutrino oscillations open  a window to explore physics beyond  the standard model, it is important to understand the underlying phenomena at the deepest level, and the domain of validity of the various calculations of oscillation probabilities and their impact on experiments. In particular, the standard approach of treating neutrino oscillations by analogy with Rabi-oscillations in a two state system (see for
example, \cite{book1,book2,book3,kayserlec,grimuslec} and references therein) while simple and intuitive, has motivated a wide ranging discussion. Deeper investigations of this basic paradigm have already raised a number of important and fundamental questions \cite{kayser,rich,nauenberg,lipkin} that are still being debated \cite{akmerev}.

A correct interpretation of the results from oscillation experiments require understanding of both the production and detection mechanisms\cite{kiers,grimusreal,giunticohe}. Neutrino detection is indirect through charged or neutral current processes and mostly through the detection of the associated charged lepton. As for the production mechanism, the neutrino state is produced by the decay of a parent particle via charged current interactions. Coherence (and decoherence) aspects of the production and detection of neutrinos \cite{giunticohe,dolgov,beuthe} and lifetime of the source\cite{stock} have been discussed,  however  only recently the recognition that the neutrino state produced by the decay of a parent particle via charged current interactions is in fact \emph{entangled} with that of the charged lepton\cite{nauenberg,goldman,glashow,nuestro,hamish,losecco,ahluwalia,patkos,aksmir} has become the focus of a reassessment of the dynamics of neutrino oscillations.

Quantum entanglement is a direct consequence of conservation laws in the production process\cite{nauenberg,goldman} which result in a correlated quantum state of the neutrino and its charged current lepton partner. As observed in refs.\cite{glashow,nuestro} in order for neutrino mass eigenstates to interfere coherently and   oscillate, the quantum state must be \emph{disentangled}: entanglement surviving for a very long time projects out states of definite energy and prevent oscillations. In a typical experiment disentanglement of the charged lepton occurs when this particle is measured, absorbed or decays, after disentanglement  the quantum state is reduced and is re-set. The full dynamics of the process of production of the entangled state, disentanglement
and further evolution to production of another charged lepton at a (far) detector was studied in a model in ref.\cite{nuestro} with a focus on long baseline experiments. Ref.\cite{patkos} studied the  \emph{free} time evolution of a disentangled wave-packet produced from pion decay including lifetime effects but without addressing the production of charged leptons by the disentangled state and detection process  .

 The results of  ref.\cite{nuestro} show that if disentanglement occurs on time scales \emph{much shorter} than the oscillation scale, namely for \emph{long baseline experiments},  the familiar result obtained from the simple quantum mechanical picture and factorization is reproduced, but also point out possible subtle consequences if the disentanglement process occurs on time scales of the same order of or longer than the oscillation time scale, with potential impact on \emph{short baseline experiments}.  This possibility is also hinted at in ref.\cite{ahluwalia}.

 In the last few years several experimental results have been accumulating that cannot be interpreted within the ``standard paradigm'' of mixing and oscillations among three ``active'' neutrinos with $\delta m^2 \simeq 10^{-4}-10^{-3}$. The early results from
 the LSND experiment\cite{lsnd} have recently been confirmed by   MiniBooNE running in antineutrino mode\cite{miniboone} both suggesting the possibility of new ``sterile'' neutrinos with $\delta m^2 \sim \mathrm{eV}^2$. More recently, a re-examination of the antineutrino flux\cite{flux1} in anticipation of the Double Chooz reactor experiment resulted in a small increase in the flux of about $3.5\%$  for reactor experiments leading to a larger deficit of $5.7\%$ suggesting a \emph{reactor anomaly}\cite{reactor}. If this deficit is the result of neutrino mixing and oscillation  with baselines $L\lesssim 10-100\,\mathrm{m}$, it requires the existence of at least one sterile neutrino with $\delta m^2 \gtrsim 1.5 \,\mathrm{eV}^2$ and mixing amplitude $\sin^2(2\theta) \simeq 0.115$\cite{reactor}. Taken together these results may be explained by models that incorporate one or more sterile neutrinos that mix with the active ones\cite{sterile1,sterile3,sterile2,giuntishort} including perhaps non-standard interactions\cite{akshwet}. Furthermore the latest analysis of the cosmic microwave background anisotropies by WMAP\cite{wmap} suggests that the effective number of   neutrino species is $N_{eff} = 4.34\pm 0.86$ and $\sum(m_\nu)<0.58\,eV$ bolstering the case for sterile neutrino(s) with $m \lesssim  \mathrm{eV}$.

 The common aspect of accelerator and reactor  anomalies is that   these are all short baseline experiments with $10\,\mathrm{m} \lesssim L \lesssim 600 \,\mathrm{m}$ and this aspect, when considered along with the potentially relevant corrections from disentanglement discussed above, motivate our study of the disentanglement and lifetime effects on short baseline oscillations. Although the effect of the muon lifetime   on neutrino oscillations has been studied in ref.\cite{grimusww} and more recently  ref.\cite{smirnov2} argued that the pion lifetime introduces decoherence  on oscillations $\nu_\mu-\nu_s$ at the near detector of the MINOS experiment, the combined effect of lifetime of the source \emph{and} disentanglement of charged leptons have not yet been discussed with regard to the distortion of the spectrum for charged lepton events at the detector in short baseline experiments.

 \vspace{3mm}

 \textbf{Goals:}
 The accumulation of experimental evidence of short baseline anomalies and the recognition that entanglement and lifetime effects may lead to corrections in the oscillation probabilities precisely in short baseline experiments motivates us to understand the impact of these effects on the appearance and disappearance probabilities, and their possible experimental implications.

 In this article we seek to understand the subtle aspects arising from quantum mechanical correlations as a result of the fact that the neutrino states produced at charged current vertices are entangled with the charged lepton partner. Measurement, absorption or decay of this charged lepton disentangles the quantum state, but the emerging neutrino state carries information on the quantum correlations in its evolution. These correlations along with intrinsic energy uncertainties associated with the lifetime of the parent particle whose decay produces the neutrinos, influence the oscillation probabilities.
 Our goal is to study these corrections in the simplest and most clear setting that allow a systematic calculation of the effects and to extract the possible impact of these effects on the experimental observables and their interpretation.

 \vspace{3mm}

 \textbf{Results:}

 \vspace{3mm}

 The dimensionless parameter that determines the impact of the lifetime of the source and the disentanglement time scale on the oscillation probabilities is
 $$ \pi L_s/L_{osc}$$
 where $L_s$ is the \emph{smaller} between the decay length of the source and the disentanglement length scale at which the charged lepton produced with the neutrino is measured, absorbed or decays and $L_{osc} \propto E_\nu /\delta m^2$ is the oscillation length.

 A detailed analysis of the production and disentanglement of the quantum states of neutrinos produced in charged current vertices reveals that the usual ``Pontecorvo'' states familiar from the simple quantum mechanical approach, are a reliable description only when $L_s \ll L_{osc}$ in which case appearance and disappearance probabilities are given by the usual expressions, but for $L_s \gg L_{osc}$ the energy uncertainties become smaller than the distance between energy eigenstates and coherence is suppressed with a concomitant suppression of the appearance probability.

 Appearance   probabilities are suppressed in short baseline experiments where both the lifetime of the source and the disentanglement scale are comparable to or a large fraction of the baseline.

 We find that at MiniBooNE the dominant source of suppression is the decay length of the pions, whereas at LSND and reactor experiments we argue that the relevant length scale   is the disentanglement distance.

 In all these cases, fits of the experimental data to the usual quantum mechanical appearance and disappearance probabilities   \emph{underestimate} both $\delta m^2$ and $\sin^2(2\theta)$. These fits are much less reliable at \emph{low neutrino energy} (for fixed $\delta m^2~,~L$) because for low energy events the ratio $L_s/L_{osc}$ is larger and the suppression of the oscillation probability is stronger.

 The corrections from lifetime and disentanglement effects on short baseline experiments are more dramatic for the mass range $\delta m^2 \sim 1\,\mathrm{eV}^2$ which is the putative mass range for sterile neutrinos that \emph{could} explain the short baseline anomalies, and for low neutrino energy.

\section{A   Model of ``Neutrino'' Oscillations}\label{sec:model}

  In order to exhibit the main results in a clear and
simple manner, we introduce a bosonic model that describes mixing,
oscillations and charged current weak interactions reliably. The complications associated with fermionic and gauge fields   are irrelevant to the
physics of mixing and oscillations, as is obviously manifest in the case of
meson mixing.

We study the model defined by the   Lagrangian density\be \mathcal{L} =
\mathcal{L}_0[W,l_\alpha]+ \mathcal{L}_0[\nu_\alpha] +
\mathcal{L}_{\rm int}[\pi,W,\l_{\alpha},\nu_\alpha] ~~;~~ \alpha=e,\mu
\label{totallag} \ee with \be {\cal L}_0[\nu] =
\frac{1}{2}\left[\partial_{\mu}\Psi^T
\partial^{\mu}\Psi  -\Psi^{T} \mathbb{M}\Psi  \right] \label{nulag}\, ,
\ee where  $\Psi$ is a flavor doublet representing the neutrinos\be
\Psi = \left(
             \begin{array}{c}
               \nu_e \\
               \nu_\mu \\
             \end{array}
           \right), \label{doublet}\ee and $\mathbb{M}$ is the mass matrix
            \be \mathbb{M} = \left(
                               \begin{array}{cc}
                                 m_{ee} & m_{e\mu} \\
                                 m_{e\mu} & m_{\mu \mu} \\
                               \end{array}
                             \right)\,. \label{massmtx}\ee
The interaction Lagrangian is similar to the charged current interaction of the standard model but explicitly includes a vertex that describes the decay of a parent particle (here the pion) into a charged lepton and its flavor neutrino $\pi \rightarrow \mu\,\nu_\mu$, namely
\be {\cal L}_{\rm int} (\vx,t) = g_{\pi}\pi(\vx,t)\mu(\vx,t)\nu_\mu(\vx,t)+g\, W(\vx,t)\Big[  {e}( \vec{x},t)\,\nu_{e}(
\vec{x},t)+  {\mu}( \vec{x},t)\,\nu_{\mu}( \vec{x},t)\Big],
\label{Interaction} \ee where $g$ plays the role of the electroweak charged current coupling, and $g_\pi$ includes the pion decay constant.  $\pi(x)$ represents the pion field or alternatively any parent particle that decays into a charged lepton and its associated neutrino,   $W(x)$
represents the vector boson,  and
$l = e,\mu$ the two charged leptons.

Obviously this simple model cannot describe CP violating effects or distinguish between neutrino vs. antineutrino modes, however our goal is to understand decoherence effects associated with lifetime of the source and disentanglement of charged leptons.

The mass
matrix is diagonalized by a unitary transformation \be
U^{-1}(\theta) \, \mathbb{M}\, U(\theta) = \left(
                                                           \begin{array}{cc}
                                                             m_1 & 0 \\
                                                             0 & m_2 \\\end{array}\right)
 ~~;~~ U(\theta) = \left(                                                                                  \begin{array}{cc}                                                                                       \cos \theta & \sin \theta \\                                                                                       -\sin\theta & \cos\theta \\                                                                                     \end{array}                                                                                   \right) . \label{massU}\ee In terms of the
  doublet of mass eigenstates,  the flavor doublet can be expressed as
  \be \left(
                                     \begin{array}{c}
                                       \nu_e \\
                                       \nu_\mu \\
                                     \end{array}
                                   \right) = U(\theta)\,\left(
                                               \begin{array}{c}
                                                 \nu_1 \\
                                                 \nu_2 \\
                                               \end{array}
                                             \right) \,.\label{masseigen}\ee
This bosonic model clearly describes charged current weak
interactions reliably as it includes all the relevant aspects of
mixing and oscillations. Furthermore the coupling $\pi\,\mu\,\nu_\mu$ allows us to study the dynamics of the production process including the lifetime of the source (  pion) within the same model.

In order to clearly separate the effects from the lifetime of the source and disentanglement time scale from the effects of wave-packet localization, this article is primarily devoted to the analysis in terms of plane waves, in section (\ref{wavepack}) we comment on the modifications from a wave packet treatment, but  postpone the full treatment with wavepackets to a more thorough  forthcoming study\cite{junme}.

We consider the case in which a neutrino and its flavor charged lepton partner are produced via the
decay of a parent particle, in this case a pion, however, the discussion and the main consequences are general, with the only difference being the associated many-particle phase space if the decay is in more than two particles. The production process corresponds to $\pi \rightarrow \mu \,\nu_\mu$ where the $\mu$ is ``observed'' or is absorbed  (or decays) at a ``disentanglement'' time scale $t_\mu$, the disentangled neutrino is detected via a charged current interaction $\nu_\mu \rightarrow W\,l $ where $l=e,\mu$ is the charged lepton.

The Wigner-Weisskopf\cite{ww,qoptics} method described in the appendix yields the entangled state that results from pion decay, the relevant part of the interaction Hamiltonian in the interaction picture is
\be H^{\pi}_I(t)= g_\pi \int d^3x ~\pi(\vx,t)\mu(\vx,t)\nu_\mu(\vx,t) \label{HIpi}\ee where the time evolution is that of free fields. The neutrino \emph{field operator}
\be \nu_\mu(\vx,t) = \cos(\theta) \nu_2(\vx,t)-\sin(\theta)\nu_1(\vx,t)\,, \label{fieldops}\ee where $\nu_{1,2}(\vx,t)$ are expanded, as usual, in  annihilation and  creation   operators of \emph{mass eigenstates} $a_{1,2}(\vp);a^\dagger_{1,2}(\vp)$ respectively, with the single particle mass eigenstates being
\be a^\dagger_{i}(\vp)|0\rangle = |\nu_{\vp,i}\rangle \label{nustates}\ee where $|0\rangle$ is the vacuum state annihilated by $a_i(\vp)$. We note that the transformation law (\ref{masseigen}) applies to the field operators not to the single particle states.

Consider that the initial state, at $t=0$ is given by   a single particle pion state described by a plane wave   with momentum $\vk$, namely
\be |\Psi(t=0) \rangle= |\pi_{\vk}\rangle \,. \label{psi0}\ee     $H^{\pi}_I$ connects the single particle pion state to the   states   $|\kappa\rangle = |\mu_{\vq},\rangle|\nu_{\vp,i}\rangle $. The transition matrix element is given by
\be \langle \mu_{\vq},\nu_{ \vp,i}|H^{\pi}_I(t)|\Pi_{\vk}\rangle \equiv \mathcal{M}_i(\vk,\vq,\vp,t) = \frac{g_{\pi}}{\sqrt{V}}\,U_{\mu\,i}~\delta_{\vk,\vp+\vq} ~\frac{e^{-i(E_{\pi}(k)-E_{\mu}(q)-\Omega_{i}(p))t}}{\sqrt{8~E_{\pi}(k) E_{\mu}(q) \Omega_{i}(p)}}\label{Mpi}\ee where $\Omega_i(p)=\sqrt{p^2+m^2_i}$ are the energies of the neutrino mass eigenstates.

We are now in position to use the results of the appendix for the time evolved state in the Schroedinger picture resulting from pion decay, it is given by eqn. (\ref{schpicstate}) with $C_A(0)=C_{\pi}(0)=1$ and the set $\kappa$ described above, we find
\be  |\Psi(t)\rangle_S =    e^{-iE_\pi(k)\,t}\,e^{-\frac{\Gamma_\pi(k)}{2}\,t}|\pi_{\vk}\rangle - \sum_{i;\vp,\vq} \mathcal{C}_i(\vk,\vq,\vp,t) ~ e^{-i(E_\mu(q)+\Omega_i(p))\,t}\,|\mu_{\vq}\rangle~|\nu_{\vp,i}\rangle  ~~;~~\vec{q}=\vk-\vec{p}   \,,\label{entstate}\ee  where the amplitudes
\be \mathcal{C}_i(\vk,\vq,\vp,t) =  \mathcal{M}_i(\vk,\vq,\vp,t=0)~ \Bigg[\frac{1-e^{-i(E_S-\Omega_i(p))t}\,e^{- \Gamma_\pi(k)\,t /2}}{E_S-\Omega_i(p)-\frac{i}{2} \Gamma_\pi(k)} \Bigg]~;~E_S= E_\pi(k)-E_\mu(q)\label{coefww} \ee
$E_{\pi}(k)$ is the fully renormalized pion energy (including the self-energy correction from the intermediate states, see appendix) and $\Gamma_{\pi}(k)= M_{\pi}\Gamma_{o}/E_\pi(k)$ where $\Gamma_o$ is the decay rate of the pion at rest. For $\Gamma_\pi =0$ the entangled  state (\ref{entstate}) is the same as that obtained in ref.\cite{nuestro} in lowest order in perturbation theory.

Although the particular form of $\mathcal{M}_i(\vk,\vq,\vp,t) $ for the bosonic theory considered  here is given by (\ref{Mpi}) the results (\ref{entstate},\ref{coefww}) are \emph{general} in terms of the transition matrix element $\mathcal{M}_i(\vk,\vq,\vp,t=0) $.

The state (\ref{entstate}) is an \emph{entangled state} of the neutrino mass eigenstates   and the muon, the entanglement is evident in that it is a \emph{sum} of product states, \emph{not a simple product state}, the amplitudes $\mathcal{C}$ are a measure of the correlation between $\nu_i$ and the charged lepton. The entanglement is a consequence of \emph{momentum conservation} since $\vq=\vk - \vp$. The ``observation'', measurement  or decay of the muon state at time $t_{\mu}$ \emph{disentangles} the neutrino state. If the muon is ``measured'' in a plane wave state with momentum $\vec{Q}$ the disentangled state is obtained by projecting the quantum state (\ref{entstate}) onto the state $|\mu_{\vec{Q}}\rangle$, namely, the disentangled neutrino
state is given by
  \be |\mathcal{V}_{\mu}(t_\mu) \rangle = \langle \mu_{\vec{Q}}|\Psi(t_{\mu})\rangle_S = - \frac{g_{\pi}e^{-iE_{\mu}(Q)t_\mu}}{\Big[4VE_\pi(k)E_\mu(Q)\Big]^\frac{1}{2}}\,\sum_{i=1,2}   \frac{ U_{\mu,i} \,e^{-i\Omega_i(p)t_\mu}}{\sqrt{2\,\Omega_i(p)}}\,F_i[k,Q,p\,;t_\mu]\,|\nu_{\vp,i}\rangle ~~;~~\vp=\vk-\vec{Q} \label{nudis}\ee where
  \be F_i[k,Q,p\,;t_\mu]= \Bigg[\frac{1-e^{-i(E_S-\Omega_i(p))t_\mu}\,e^{- \Gamma_\pi(k)\,t_\mu/2}}{E_S-\Omega_i(p)-\frac{i}{2} \Gamma_\pi(k)} \Bigg]~;~E_S= E_\pi(k)-E_\mu(Q) \,,\label{Fi}\ee The functions $F_i$ encode the information of production of the entangled charged-lepton-neutrino pair and the measurement of the charged lepton, it features both time scales: the lifetime of the source and the disentanglement time scale $t_\mu$.

  The number of muons of momentum $\vec{Q}$ detected at $t_\mu$ is given by
  \be \mathcal{N}_\mu(\vec{Q},t_\mu) \equiv (2\pi)^3\frac{d^6 N_\mu}{d^3xd^3\vec{Q}} = \langle \Psi(t_\mu )|a^\dagger_{\mu}(\vec{Q})a_\mu(\vec{Q})|\Psi(t_\mu )\rangle = \langle \mathcal{V}_\mu(t_\mu)|\mathcal{V}_\mu(t_\mu)\rangle \,,\label{muonnum}\ee where $a^\dagger_\mu,a_\mu$ are creation and annihilation operators for muons. Thus the normalization of the disentangled neutrino state is completely determined by the number density of muons detected at $t_\mu$.

  We find
  \be \mathcal{N}_\mu(\vec{Q},t_\mu) = \frac{g^2_\pi}{\Big[4VE_\pi(k)E_\mu(Q)\Big]}\,\Bigg[\frac{\cos^2(\theta)}{2\,\Omega_2(p)}\big|F_2(k,Q,p,t_\mu)\big|^2+
   \frac{\sin^2(\theta)}{2\,\Omega_1(p)}\big|F_1(k,Q,p,t_\mu)\big|^2\Bigg]\,.\label{numnu}\ee

   This expression becomes familiar from the following analysis: the functions
   \be \big|F_i(k,Q,p,t_\mu)\big|^2 =  \frac{ \Big(1-e^{-\frac{\Gamma_\pi(k)}{2}\,t_\mu}\Big)^2+ 4\,e^{-\frac{\Gamma_\pi(k)}{2}\,t_\mu}\,\sin^2\Big[ (E_S-\Omega_i(p))\frac{t_\mu}{2}\Big] }{\Big(E_S-\Omega_i(p))\Big)^2+\frac{\Gamma^2_\pi(k)}{4}}  \label{funis}\ee are strongly peaked at $E_S = E_\pi(k)-E_\mu(Q) = \Omega_i(p)$ with width determined by the \emph{largest} of $\Gamma_\pi(k)~;~2\pi/t_\mu$ becoming proportional to energy conserving delta functions in the limit when these become very small.    This is clearly seen in two relevant limits:

   \vspace{3mm}

   \textbf{i):} the narrow width limit with $\Gamma_\pi(k)\, t_\mu \ll 1$ but large $t_\mu $ where
   \be \big|F_i(k,Q,p,t_\mu)\big|^2 \simeq  \frac{  4 \,\sin^2\Big[ (E_S-\Omega_i(p))\frac{t_\mu}{2}\Big] }{\Big(E_S-\Omega_i(p))\Big)^2 }  \,, \label{smalga}\ee This limit corresponds to a long disentanglement time scale but $t_\mu \ll T_\pi(k)$ where $T_\pi(k)=1/\Gamma_\pi(k)$ is the pion lifetime in the laboratory frame. This function is strongly peaked at $E_S = \Omega_i(p)$ with maximum height $t^2_\mu$ and width $2\pi/t_\mu$ which is the \emph{largest} of $\Gamma_\pi(k)$ and $2\pi /t_\mu$ for this case. As $t_\mu$ becomes very large,
   \be \big|F_i(k,Q,p,t_\mu)\big|^2 \simeq   2\pi \, t_\mu \,\delta\big(E_S-\Omega_i(p)\big) \,. \label{lont}\ee

   \vspace{3mm}

   \textbf{ii:)}   the opposite limit,  for $\Gamma_\pi(k)\,t_\mu \gg 1$, in which the disentanglement time scale $t_\mu$ is much longer than $T_\pi(k)$, where
   \be \big|F_i(k,Q,p,t_\mu)\big|^2 \simeq  \frac{ 1  }{\Big(E_S-\Omega_i(p) \Big)^2+\frac{\Gamma^2_\pi(k)}{4}} \,. \label{larga}\ee The function is strongly peaked at $E_S =\Omega_i(p)$ of height $4/\Gamma^2_\pi(k)$ and width $\Gamma_\pi(k)/2$ which is the \emph{largest} of $\Gamma_\pi~;~2\pi/t_\mu$ in this case. In the narrow width limit
   \be \big|F_i(k,Q,p,t_\mu)\big|^2\simeq 2\pi T_\pi(k) \,\delta\big(E_S-\Omega_i(p)\big)\,. \label{narrow}\ee

   It is clear from this discussion that $|F_i|^2$ describe \emph{approximate} energy conservation at the production vertex, approximate because the finite disentanglement time scale $t_\mu$ and/or the pion lifetime $T_\pi(k)$ broaden the energy conserving delta functions with an energy resolution determined by the width which is the \emph{largest} of  $  2\pi/t_\mu$ or $2\pi/T_\pi(k)$ respectively, namely  the \emph{shortest} time scale.

   For the general form (\ref{funis}), a straightforward integration yields
   \be \int^\infty_{-\infty} |F_i|^2 dE_S = 2\pi T_\pi(k) \Big[1-e^{-\Gamma_\pi(k)\,t_\mu}\Big] \,, \label{integF}\ee  therefore, \emph{assuming} that the energy distribution is very sharply peaked at $E_s = \Omega_i(p)$,  we can approximate
   \be \big|F_i(k,Q,p,t_\mu)\big|^2     \simeq 2\pi T_\pi(k) \Big[1-e^{-\Gamma_\pi(k)\,t_\mu}\Big] \,\delta\big(E_S-\Omega_i(p)\big)\,. \label{aprointF}\ee

     In this  approximation   the total number of muons measured at the disentanglement time scale is
   \be N_\mu = \int d^3x \int \frac{d^3Q}{(2\pi)^3} ~\mathcal{N}_\mu(\vec{Q},t_\mu) \simeq  \Gamma_{\pi}(k)\,T_\pi(k) \,\Big[1-e^{-\Gamma_\pi(k)\,t_\mu}\Big] \label{totnummu}\ee where
    \be \Gamma_\pi(k) = \cos^2(\theta)\, \Gamma_{\pi\rightarrow \mu\,\nu_2}(k)+ \sin^2(\theta) \Gamma_{\pi\rightarrow \mu\,\nu_1}(k) \label{gammatot} \ee is the \emph{total pion decay rate} (in this simple model) and
    \be  \Gamma_{\pi\rightarrow \mu\,\nu_i}(k) = \frac{g^2_\pi}{32\,\pi^2\,E_\pi(k)}\int \frac{d^3Q }{E_\mu(Q)\,\Omega_i(p)}~ \delta\big( E_\pi(k)-E_\mu(Q)-\Omega_i(p) \big) ~~;~~\vec{p}=\vk-\vec{Q} \label{widths} \ee are the partial widths for pion decay into the neutrino \emph{mass eigenstates}. Although this result applies to the simple model considered here, clearly it is general and conceptually correct: the particle decays into \emph{mass eigenstates}  which are the correct eigenstates of the unperturbed Hamiltonian, with the probabilities determined by $\cos^2(\theta)~,~\sin^2(\theta)$ respectively.

   In what follows, we consider ultrarelativistic and nearly degenerate neutrinos and write
   \be \Omega_1(p) \simeq E_\nu(p) - \Delta(p)~~;~~
  \Omega_2(p) \simeq E_\nu(p) + \Delta(p), \label{URO}\ee where
   \be   E_\nu(p)  = \left[ p^2 +\frac{m^2_1+m^2_2}{2}\right]^\frac{1}{2} ~~;~~ \Delta(p)= \frac{\delta m^2}{4E_\nu(p)  }
   ~~;~~\delta m^2 = m^2_2-m^2_1,  \label{diffs}\ee and take
   $\Delta(p) \ll \oO(p)$ as is the experimentally relevant case. For $\Delta/\oO \ll 1$ the energy conserving $ \delta\big( E_\pi(k)-E_\mu(Q)-\Omega_i(p) \big)$ in (\ref{widths}) may be replaced by $ \delta\big( E_\pi(k)-E_\mu(Q)-\oO(p) \big)$ in the integral, because for the experimental range $\delta m^2 \lesssim 1\,\mathrm{eV}^2~;~E_\nu  \gtrsim 1\,\mathrm{MeV}$ the relative error incurred $\propto (\Delta/\oO) \lesssim 10^{-12}$ is \emph{much} smaller than typical experimental resolution.

The detection or measurement of the muon at time $t_\mu$ re-sets the quantum state to $|\mathcal{V}_\mu(t_u)\rangle$, upon further evolution in time   this disentangled state evolves into
\be |\mathcal{V}_\mu(t) \rangle = e^{-iH_0 t}\,\mathcal{U}(t,t_\mu)\,e^{iH_0 t_\mu}\,|\mathcal{V}_\mu(t_\mu) \rangle \,, \label{fulltimeevol}\ee where
\be \mathcal{U}(t,t_\mu)= T \Big(e^{i\int^t_{t_\mu}dt'\int d^3x ~ \mathcal{L}_{int}(\vec{x},t') }\Big) \label{Uop}\ee is the time evolution operator in the interaction picture with boundary condition $\mathcal{U}(t_\mu,t_\mu)=1$.

The usual ``Pontecorvo'' quantum state  familiar in the literature are simple linear superpositions $|\nu_l\rangle = \sum_i U_{li}|\nu_i\rangle$, where $U_{li} $ are the elements of the mixing matrix (\ref{massU}), and the corresponding muon neutrino state at any time $t$  is
\be |\nu_\mu\rangle (t)= \cos(\theta)\,e^{-i\Omega_2(p)\,t}\,|\nu_2\rangle -\sin(\theta)\,e^{-i\Omega_1(p)\,t}\,|\nu_1\rangle \,.\label{pontemuont}\ee  Instead, the corresponding quantum state evolved \emph{freely} in time from $t_\mu$ up to $t$ from the \emph{disentangled} state $|\mathcal{V}_\mu(t_u)\rangle$, is obtained from (\ref{fulltimeevol}) by setting $\mathcal{U}(t,t_\mu)= 1$, it is given by
\bea |{\mathcal{V}}_\mu(t)\rangle  =    e^{-iH_0(t-t_\mu)}  |{\mathcal{V}}_\mu(t_\mu)\rangle & = &  N\Bigg[ {\cos(\theta)} ~F_2(k,Q,p,t_\mu)~ e^{-i\Omega_2(p)\,t}~|\nu_{\vp,2}\rangle\nonumber \\ & - &
    {\sin(\theta)} ~F_1(k,Q,p,t_\mu)~e^{-i\Omega_1(p)\,t}~|\nu_{\vp,1}\rangle \Bigg] \label{timevolv}\eea where the prefactor $N$ can be read off (\ref{nudis}) and    we have neglected terms of $\mathcal{O}\big(\Delta(p)/\oO(p)\big)^2$ thereby approximating $\sqrt{2~\Omega_i(p)}\simeq \sqrt{2~\oO(p)}$ in (\ref{nudis}) including this factor in   $N$. Consequently we also
    replace $\Omega_{1,2}(p)\rightarrow \oO(p)$ in (\ref{numnu}).

    Obviously the usual ``Pontecorvo'' states emerge  up to the overall normalization factor  \emph{if} $|F_1|=|F_2|$ since time independent phases can be absorbed in the definition of the mass eigenstates.  The conditions under which this equality is fulfilled is analyzed below.

    Following   the familiar quantum mechanical approach to obtain the survival probability we find
    \bea \mathcal{P}_{\nu_\mu \rightarrow \nu_\mu}(t;t_\mu) =   \big|\langle {\mathcal{V}}_\mu(t_\mu)|{\mathcal{V}}_\mu(t)\rangle\big|^2 & = &  |N|^4  \Bigg\{\Big[\cos^2(\theta)|F_2|^2+\sin^2(\theta)|F_1|^2\Big]^2 \nonumber \\ & - &   |F_2|^2~ |F_1|^2 \,\sin^2(2\theta)\,\sin^2\Big[ \frac{\delta m^2}{4E_\nu(p)  }(t-t_\mu)\Big] \Bigg\}\label{diss}\eea

     Obviously if $|F_1|=|F_2|$ there is agreement with the usual result from Pontecorvo states up to an overall normalization.

    We note that invoking the approximation (\ref{aprointF}) for $|F_i|^2$ yields $|F_1F_2|^2=0$ as the product of delta functions vanishes for $\Omega_1\neq \Omega_2$ thereby leading to the \emph{hasty} conclusion that coherence is completely suppressed, however (\ref{aprointF}) is an approximation that neglects the fact that $|F_i|^2$ are not sharp distributions but broadened with typical widths $\Gamma_\pi$ or $2\pi/t_\mu$. Thus an assessment of the   coherence leading to oscillations and interference requires a careful and detailed  examination of the product $|F_1F_2|^2$.

    It proves convenient to use  (\ref{Fi},\ref{URO}) and write
    \be \big|F_1(k,Q,p;t_\mu)\big|^2 = \frac{ \Big(1-e^{-\frac{\Gamma_\pi(k)}{2}\,t_\mu}\Big)^2+ 4\,e^{-\frac{\Gamma_\pi(k)}{2}\,t_\mu}\,\sin^2\Big[ (\mathcal{E}_S+\Delta(p))\frac{t_\mu}{2}\Big] }{\Big(\mathcal{E}_S+\Delta(p)\Big)^2+\frac{\Gamma^2_\pi(k)}{4}}   \label{modf12}\ee
    \be \big|F_2(k,Q,p;t_\mu)\big|^2 = \frac{ \Big(1-e^{-\frac{\Gamma_\pi(k)}{2}\,t_\mu}\Big)^2+ 4\,e^{-\frac{\Gamma_\pi(k)}{2}\,t_\mu}\,\sin^2\Big[ (\mathcal{E}_S-\Delta(p))\frac{t_\mu}{2}\Big] }{\Big(\mathcal{E}_S-\Delta(p)\Big)^2+\frac{\Gamma^2_\pi(k)}{4}}   \label{modf22}\ee where
    \be \mathcal{E}_S=E_S-\oO(p)\,. \label{maE}\ee

    The limits studied above clarify the impact of the width of the energy distribution,

    \begin{itemize}
    \item{ \textbf{i):} $\Gamma_\pi(k)t_\mu \gg 1$. In this case
    \be \big|F_1(k,Q,p;t_\mu)\big|^2 \simeq 2\,T_\pi(k)~ \frac{\Big(\frac{\Gamma_\pi(k)}{2}\Big)}{\Big(\mathcal{E}_S+\Delta(p)\Big)^2+\frac{\Gamma^2_\pi(k)}{4} } \label{1Gtgg1} \ee
     \be \big|F_2(k,Q,p;t_\mu)\big|^2 \simeq 2\,T_\pi(k)~ \frac{\Big(\frac{\Gamma_\pi(k)}{2}\Big)}{\Big(\mathcal{E}_S-\Delta(p)\Big)^2+\frac{\Gamma^2_\pi(k)}{4} } \label{2Gtgg1} \ee
     where $T_\pi(k) = 1/\Gamma_\pi(k)$ is the pion lifetime in the laboratory frame. Each Lorentzian is peaked at $\mathcal{E}_S = \pm \Delta(p)$ respectively with a height $\propto T^2_\pi(k)$ and width $\propto \Gamma_\pi(k)$, their product   is depicted in fig. (\ref{fig:fig1}) for $\Gamma_\pi(k) > \Delta(p)$ and $\Gamma_\pi(k) \ll \Delta(p)$ respectively. For $\Gamma_\pi(k) \gg \Delta(p)$ the product is similar to one Lorentzian peaked at $\mathcal{E}_S \sim 0$ because the width of the individual Lorentzians ($\simeq \Gamma_\pi(k)$) is larger than their separation ($\simeq \Delta(p)$), therefore the lifetime of the source introduces a large energy uncertainty that cannot resolve between the nearly degenerate energy eigenstates and \emph{blurs} the individual peaks under one broad peak. In this case $|F_1|^2 \simeq |F_2|^2$ and  the disentangled state $|\mathcal{V}_\mu(t)\rangle$ is proportional to the corresponding Pontecorvo state.

      On the other hand if  $\Gamma_\pi(k) \ll \Delta(p)$ the product is a double peaked distribution with peaks at $\mathcal{E}_S \simeq \pm \Delta(p)$ of widths $\simeq \Gamma_\pi(k)$ but with height $1/\Big(\Delta^2(p)\Gamma^2_\pi(k)\Big)\ll 1/\Gamma^4_\pi(k)$\footnote{This suppression is not manifest in fig. (\ref{fig:fig1}) because we have chosen $\Delta$ as the overall scale for presentation purposes.}  as is the case for $|F_{1,2}|^4$ which are the terms that do not feature oscillations.  In this case the distance between the peaks $\simeq \Delta(p)$ is much larger than the width of the individual peaks $\simeq \Gamma_{\pi}(k)$ and the energy uncertainty $\sim \Gamma_\pi(k)$ is small enough that the individual mass eigenstates are resolved.

       This can be seen more efficiently from the identity
      \be |F_1|^2\,|F_2|^2 = \frac{1}{\mathcal{E}^2_S+\Delta^2(p) +\frac{\Gamma^2_\pi(k)}{4}} ~ \frac{1}{2}\Big[|F_1|^2+|F_2|^2\Big]\simeq  \frac{1}{2\,\Delta^2(p) +\frac{\Gamma^2_\pi(k)}{4}} ~ \frac{1}{2}\Big[|F_1|^2+|F_2|^2\Big] \label{idprods}\ee where the last approximate equality follows from the fact that $|F_{1,2}|^2$ are strongly peaked at $\mathcal{E}_S \mp \Delta(p)$.

     \begin{figure}[h!]
\begin{center}
\includegraphics[height=3in,width=3in,keepaspectratio=true]{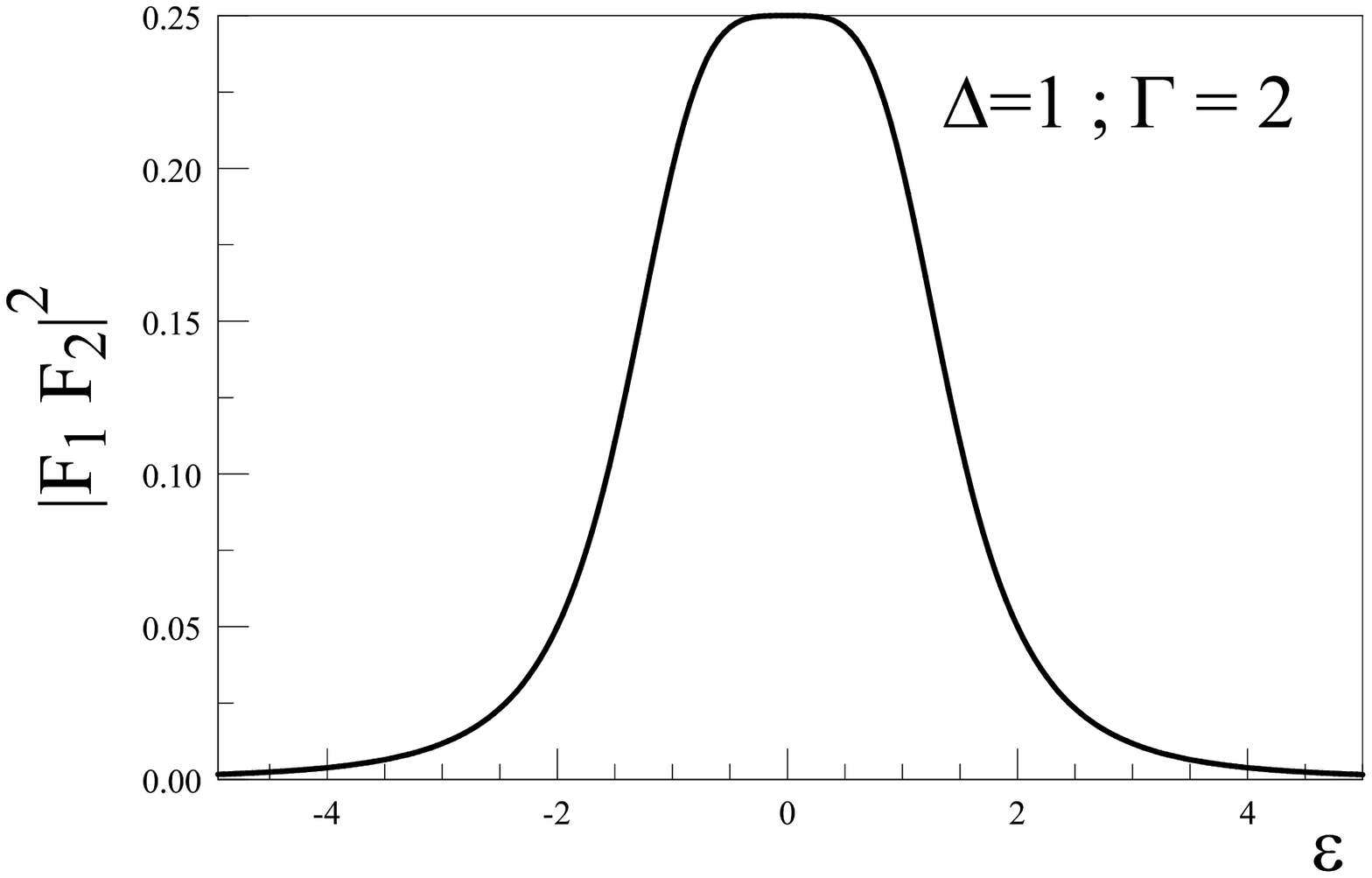}
\includegraphics[height=3in,width=3in,keepaspectratio=true]{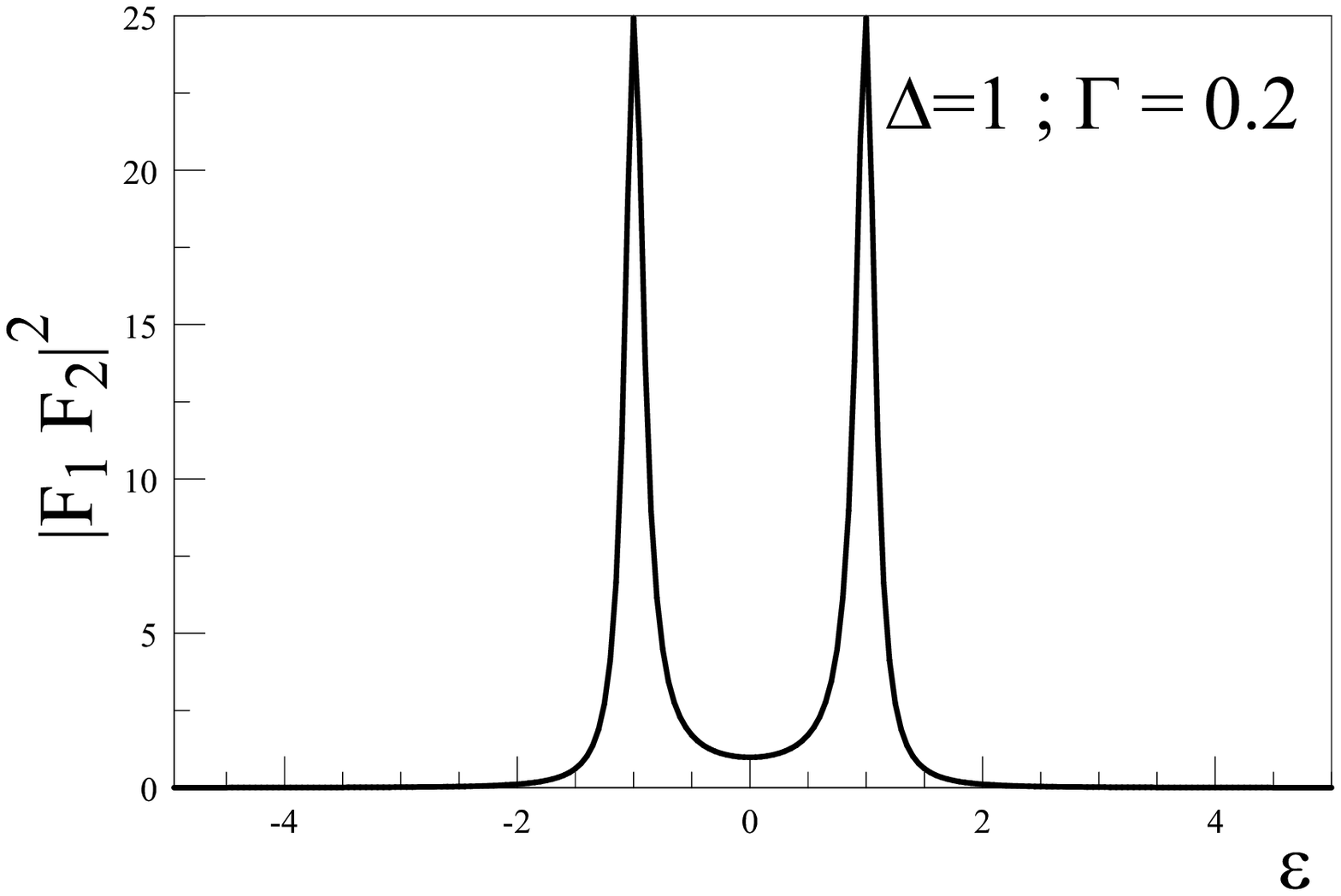}
\caption{The product $\big|F_1(k,Q,p;t_\mu)F_2(k,Q,p;t_\mu)\big|^2$ vs. $\mathcal{E}_S$ in units of $\Delta$ for $\Gamma_\pi(k) =2,0.2$ respectively for $\Gamma_\pi(k)t_\mu\gg 1$. } \label{fig:fig1}
\end{center}
\end{figure}

   Therefore, the conclusion is that when $\Delta(p) \gg \Gamma_\pi(k)$ and $t_\mu \gg T_\pi$  it follows that  $ |F_1|^2|F_2|^2 \ll |F_{1,2}|^4$ and coherence and oscillations are \emph{suppressed}.

 }

\item{\textbf{ii):} $\Gamma_\pi(k)t_\mu \ll 1$. In this case
 \be \big|F_1(k,Q,p;t_\mu)\big|^2 \simeq \frac{ \Big(\frac{\Gamma_\pi(k)}{2}\,t_\mu\Big)^2+ 4 \,\sin^2\Big[ (\mathcal{E}_S+\Delta(p))\frac{t_\mu}{2}\Big] }{\Big(\mathcal{E}_S+\Delta(p)\Big)^2+\frac{\Gamma^2_\pi(k)}{4}}   \label{modf12small}\ee
    \be \big|F_2(k,Q,p;t_\mu)\big|^2 = \frac{ \Big( \frac{\Gamma_\pi(k)}{2}\,t_\mu\Big)^2+ 4\, \sin^2\Big[ (\mathcal{E}_S-\Delta(p))\frac{t_\mu}{2}\Big] }{\Big(\mathcal{E}_S-\Delta(p)\Big)^2+\frac{\Gamma^2_\pi(k)}{4}} \,.  \label{modf22small}\ee These are sharply peaked distributions that feature maxima at $\mathcal{E} = \mp\Delta(p) $ respectively with heights $t^2_\mu$ and widths $\simeq 2\pi/t_\mu$,   displayed in fig. (\ref{fig:fig2}) for $ \Delta(p)=1;~\Gamma_\pi(k) =0.01;~ t_\mu = 2,20 $. If $2\pi/t_\mu > \Delta(p)$ the two peaks are blurred into one broad peak, whereas if $2\pi/t_\mu \ll \Delta(p)$ the peaks are separated, this is a manifestation of the separation of mass eigenstates in real time as a consequence of the energy-time uncertainty, nevertheless, the product $|F_1F_2|^2$ is not only \emph{not} vanishing but with large support at the individual peaks.

     \begin{figure}[h!]
\begin{center}
\includegraphics[height=3in,width=3in,keepaspectratio=true]{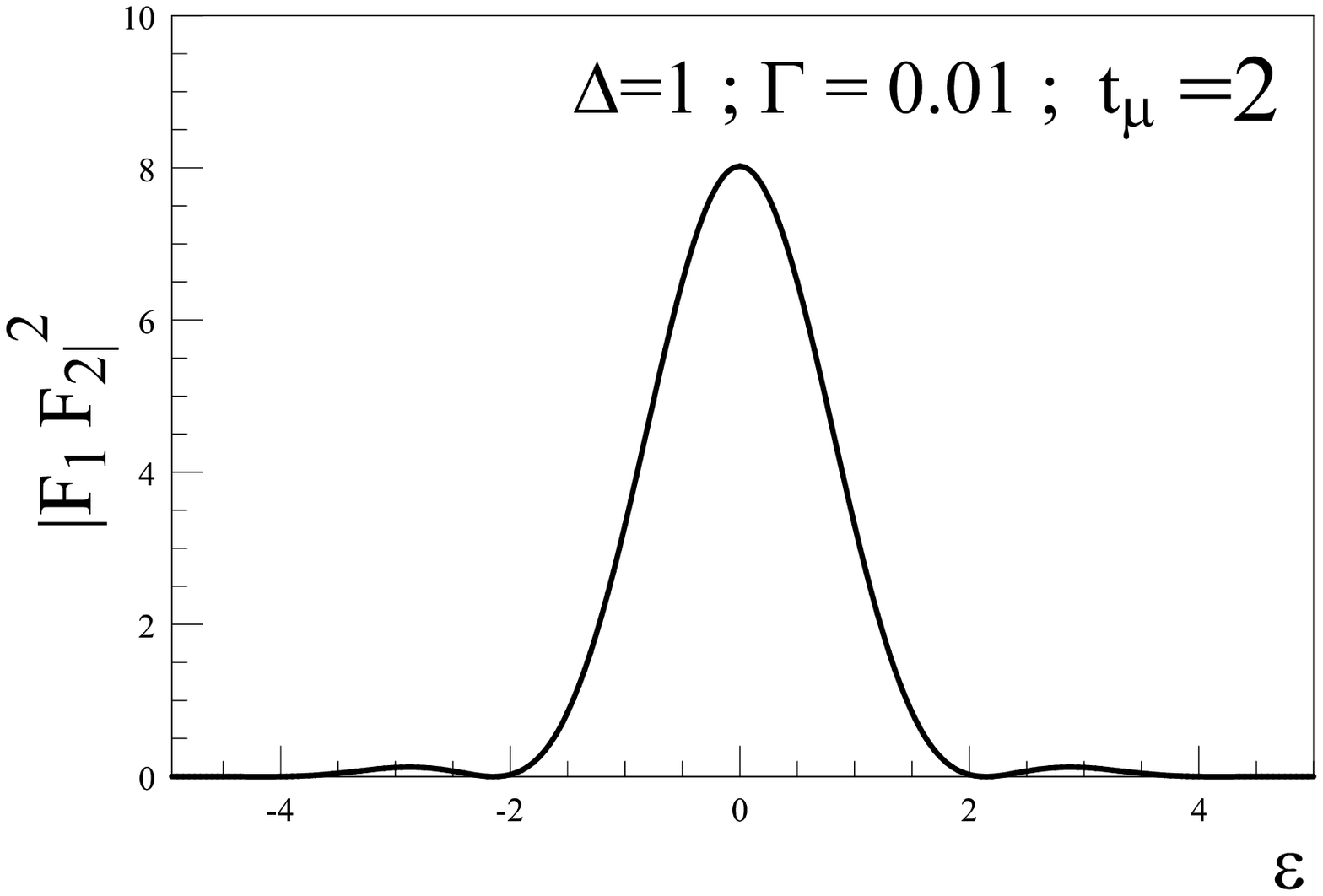}
\includegraphics[height=3in,width=3in,keepaspectratio=true]{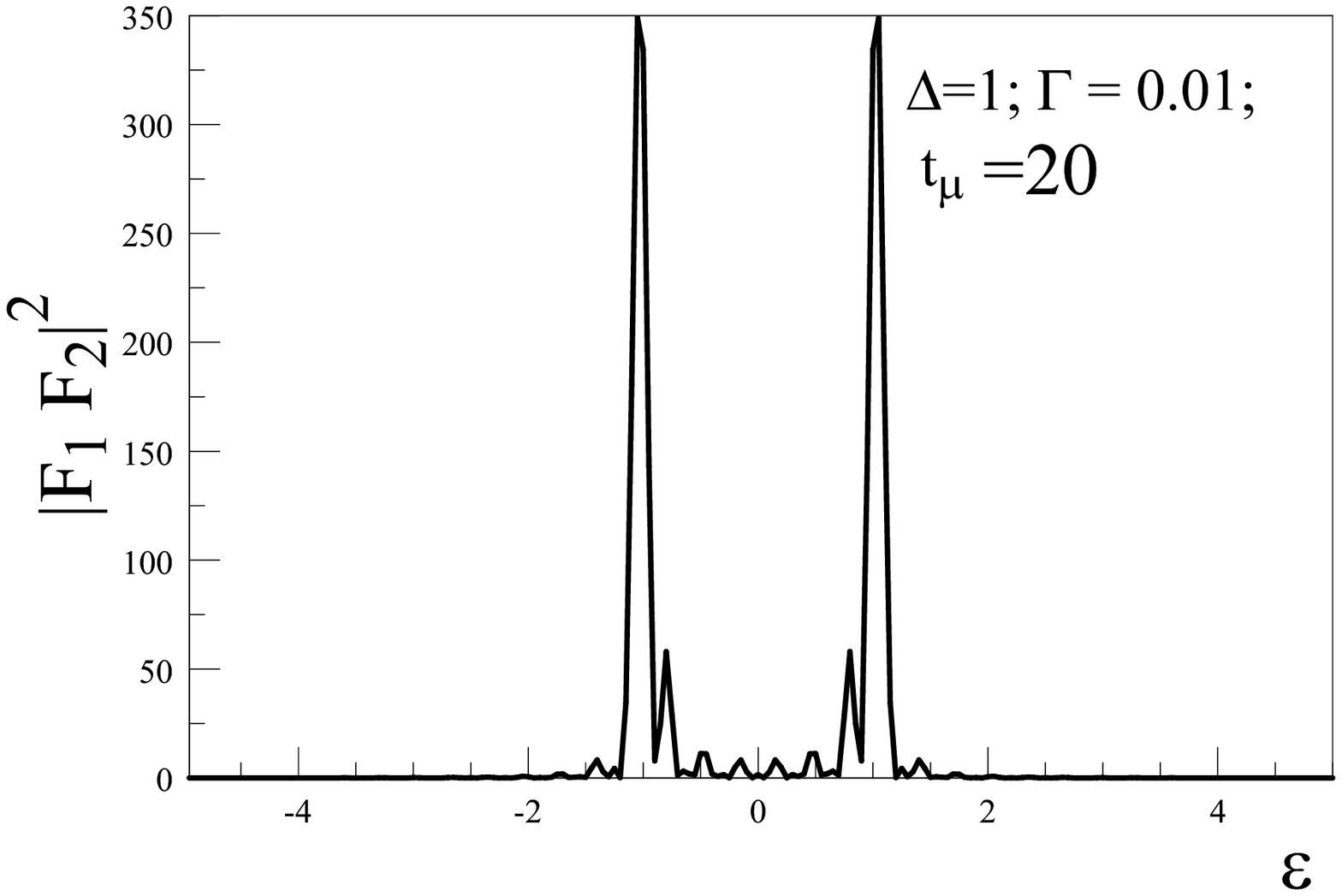}
\caption{The product $\big|F_1(k,Q,p;t_\mu)F_2(k,Q,p;t_\mu)\big|^2$ vs. $\mathcal{E}_S$ in units of $\Delta$ for $\Gamma_\pi(k) =0.01$   for $\Gamma_\pi(k)t_\mu\ll 1$ for $t_\mu = 2,20$ respectively. } \label{fig:fig2}
\end{center}
\end{figure}

An important case is when $\Gamma_\pi=0$, namely a stationary source, which will be important in the discussion below. In this case for large $t_\mu$
\be |F_{1,2}|^2 = 4\,\frac{  \sin^2\Big[ (\mathcal{E}_S\pm\Delta(p))\frac{t_\mu}{2}\Big] }{\Big(\mathcal{E}_S\pm\Delta(p)\Big)^2 } \simeq 2\pi\,t_\mu \,\delta(\mathcal{E}_S \pm \Delta(p) ) \label{Gzero}\ee whereas
\bea |F_1 F_2|    & =& 4\,\frac{\sin[(\mathcal{E}_S-\Delta)\frac{t_\mu}{2}]}{(\mathcal{E}_S-\Delta)}\,
 \frac{\sin[(\mathcal{E}_S+\Delta)\frac{t_\mu}{2}]}{(\mathcal{E}_S+\Delta)} \nonumber \\ & = & \frac{\sin(\Delta\,t_\mu)}{\Delta}\Bigg[\frac{\sin[(\mathcal{E}_S-\Delta)t_\mu]}{(\mathcal{E}_S-\Delta)}+
 \frac{\sin[(\mathcal{E}_S+\Delta)t_\mu]}{(\mathcal{E}_S+\Delta)}\Bigg]   \nonumber\\ & & + \, 2\frac{\cos(\Delta\,t_\mu)}{\Delta}\Bigg[\frac{\sin^2[(\mathcal{E}_S-\Delta)\frac{t_\mu}{2}]}{(\mathcal{E}_S-\Delta)}- \frac{\sin^2[(\mathcal{E}_S+\Delta)\frac{t_\mu}{2}]}{(\mathcal{E}_S+\Delta)}\Bigg]\,. \label{prodagf1f2}\eea In the long time limit, in   the first line we can replace
\be \frac{\sin[(\mathcal{E}_S\pm\Delta)t_\mu]} {(\mathcal{E}_S\pm\Delta)} \simeq \pi \delta(\mathcal{E}_S\pm\Delta)\,,\label{delfun}\ee
  whereas the second term is subleading in the limit $t_\mu \rightarrow \infty;\Delta \rightarrow 0$, therefore
 $|F_1F_2| \simeq \sin(\Delta(p)\,t_\mu)/(\Delta(p)\,t_\mu) ~|F_{1,2}|^2$, namely in eqn. (\ref{diss}) the interference term is of the same order as the direct terms for $\Delta(p)\, t_\mu \ll 1$ but suppressed   for $\Delta(p)\, t_\mu \gg 1$.

     }

    \end{itemize}

We highlight that $\Delta(p) = \pi/t_{osc}$ where $t_{osc}$ is the oscillation time scale, therefore the conclusion from the analysis above is  that $F_1 \simeq F_2$  when the smallest of $T_\pi(k),t_\mu \ll t_{osc}$  and the time uncertainty due to either the lifetime of the source or the disentanglement time scale leads to an energy uncertainty $\simeq \Gamma_\pi(k), 2\pi/t_\mu$ that cannot distinguish between the mass eigenstates, and the resulting quantum state is similar to a ``Pontecorvo'' state namely a coherent superposition of mass eigenstates leading to oscillations   with the familiar quantum mechanical survival probability.

 On the other hand if $T_\pi(k),t_\mu \gtrsim t_{osc}$ the mass eigenstates separate in time as the energy uncertainty becomes smaller than the distance between the broadened ``delta functions'', the coherence  and interference between the mass eigenstates is \emph{suppressed} by this separation and the disentangled state is not of the form of a ``Pontecorvo state''.

 The energy uncertainty is determined by the \emph{smallest} of the lifetime of the source and the disentanglement time scale. When this uncertainty is larger than the energy separation between neutrino mass eigenstates, the disentangled quantum states is the ``Pontecorvo'' state, but if the uncertainty is much smaller than the energy separation coherence and oscillations are suppressed.

\emph{Therefore entanglement  during a time scale of the order of or longer than the oscillation time scale   suppress    coherence and oscillations.}

    These are some of the main results of this article.

These results were obtained from  the free evolution of the disentangled state, however, what is needed is an assessment of how these considerations are manifest in the energy spectrum of the charged leptons that are measured at the detector.

\section{Charged lepton detection:}

To obtain the  transition probability to a final state with a charged lepton, which is
the state finally detected, we must go to first order in the charged current interaction in (\ref{Interaction}), evolving the disentangled state $ | {\mathcal{V}}_\mu(t_\mu)\rangle$ up to time
$t_D$ at which the charged lepton is detected. In first order in the charged current interaction we find from (\ref{fulltimeevol})
\be   | {\mathcal{V}}_\mu(t_D)\rangle^{(1)}  = e^{-iH_0\,t_D}\, (-i g)  \,\sum_l\,\int^{t_D}_{t_\mu} dt'\int d^3x\, U_{l\,i} W(\vx,t')\,l(\vx,t')\,\nu_i(\vx,t')~e^{iH_0\,t_\mu}~| {\mathcal{V}}_\mu(t_\mu)\rangle \,. \label{1stord}\ee The contributions that yield a $W$ and a charged lepton $l $ in the final state correspond  to annihilating
the neutrino states from $| {\mathcal{V}}_\mu(t_\mu)\rangle$ and creating the final $W,l $. A straightforward calculation yields,

\bea    | {\mathcal{V}}_\mu(t_D)\rangle^{(1)}  & = &  ~(-ig\,g_\pi)\,e^{-iE_\mu(Q)t_\mu}\,\sum_l \sum_{\vec{q}_l} V ~\Pi_l \Bigg\{ U_{l\, 2}~
\frac{\cos(\theta)}{2\Omega_2(p)}F_2[k,Q,p,t_\mu]~G_2[p,q_l,t_D,t_\mu] \nonumber \\ & - &  U_{l\,  1}~
\frac{\sin(\theta)}{2\Omega_1(p)}F_1[k,Q,p,t_\mu]~G_1[p,q_l,t_D,t_\mu]\Bigg\} ~e^{-iE_D\,t_D}\, |W_{\vec{k}_w}\rangle\,|l_{\vec{q}_l}\rangle \label{finstate} \eea where
\be \vec{p}=\vec{k}-\vec{Q} ~;~  \vec{k}_w=\vp-\vec{q}_l~;~ E_D=E_W(k_w)+E_{l }(q_l)~;~E_S=E_\pi(k)-E_\mu(Q)\, \label{conservation}\ee
$$\Pi_l  = \Bigg[V^4\,2E_\pi(k)\,2E_\mu(Q)\,2E_W(k_w)\,2E_{l}(q_l)\Bigg]^{-1/2} $$
and
\be G_i[p,q_l,t_D,t_\mu] = \int^{t_D}_{t_\mu} dt'~ e^{i(E_D-\Omega_i(p))t'} = e^{i(E_D-\Omega_i(p))(t_D+t_\mu)/2}~~ \frac{ \sin\Big[(E_D-\Omega_i)(t_D-t_\mu)/2\Big]}{\Big[(E_D-\Omega_i)/2\Big]}  \label{Gis}\ee The functions $F_{1,2}$ and $G_{1,2}$ determine \emph{approximate} energy conservation at the production ($F_{1,2}$) and detection ($G_{1,2}$) vertices respectively.

The relevant observable is the energy distribution of the charged leptons measured in the detector, namely
\be \mathcal{N}_l(\vq_l,t_D) = ~ {}^{(1)}\langle\mathcal{V}_\mu(t_D)|a^\dagger_l(\vq_l)a_l(\vq_l)|\mathcal{V}_\mu(t_D)\rangle^{(1)} \label{Nl}\ee using $\Omega_i(p)\approx \oO(p)$ in the denominators,  we find
\be \mathcal{N}_\mu(\vq_\mu,t_D) = \Bigg[\frac{gg_\pi\,V\,\Pi_\mu  }{2~\oO(p)}\Bigg]^2 ~\Bigg\{\cos^4(\theta)\,|F_2|^2\,|G_2|^2+\sin^4(\theta)\,|F_1|^2\,|G_1|^2+\frac{1}{2}\sin^2(2\theta)\,\mathrm{Re}
\Big[ F_2\,F^*_1 \,G_2\,G^*_1
\Big] \Bigg\} \label{mudec}\ee

\be \mathcal{N}_e(\vq_e,t_D) = \Bigg[\frac{gg_\pi\,V\,\Pi_e  }{2~\oO(p)}\Bigg]^2 \,\frac{1}{2}\sin^2(2\theta)~ ~\Bigg\{\frac{1}{2} \,|F_2|^2\,|G_2|^2+\frac{1}{2} \,|F_1|^2\,|G_1|^2-\,\mathrm{Re}
\Big[ F_2\,F^*_1 \,G_2\,G^*_1
\Big] \Bigg\}\,. \label{eldec}\ee

The term $\mathrm{Re} \Big[ F_2\,F^*_1 \,G_2\,G^*_1 \Big]$ describes the interference between the mass eigenstates that include the initial correlation in the entangled quantum state (\ref{entstate}).

As a guide, we note that \emph{if} $|F_1|=|F_2|$ \emph{and} $|G_1|=|G_2|$  then one would find
\be \mathcal{N}_\mu(\vq_\mu,t_D) = \Bigg[\frac{gg_\pi\,V\,\Pi_\mu  }{2~\oO(p)}\Bigg]^2~|F_2|^2\,|G_2|^2 ~\Bigg\{1- \sin^2(2\theta)\,\sin^2\Big[\frac{\Phi}{2}\Big] \Bigg\} \label{mudeceq}\ee

\be \mathcal{N}_e(\vq_e,t_D) = \Bigg[\frac{gg_\pi\,V\,\Pi_e }{2~\oO(p)}\Bigg]^2 ~|F_2|^2\,|G_2|^2 ~ \,\Bigg\{\sin^2(2\theta)\,\sin^2\Big[\frac{\Phi}{2}\Big]\Bigg\} \label{eldeceq}\ee where $\Phi$ is the total phase in the product inside the real part in (\ref{mudec},\ref{eldec}). Clearly the terms inside the brackets in (\ref{mudeceq},\ref{eldeceq}) are the disappearance $\mathcal{P}_{\nu_\mu \rightarrow \nu_\mu}$ and appearance $\mathcal{P}_{\nu_\mu \rightarrow \nu_e}$ probabilities associated with Pontecorvo states respectively.

The general case is obtained by replacing $F_{1,2}$ by (\ref{Fi}) and $G_{1,2}$ by (\ref{Gis})in the interference term
\be I= \mathrm{Re}
\Big[ F_2\,F^*_1 \,G_2\,G^*_1
\Big] \,.\label{interf}\ee

Before we study this interference term in detail, let us consider again the direct terms $|F_{1,2}|^2$ and $|G_{1,2}|^2$. The functions $|F_{1,2}|^2$ are sharply peaked at $E_S = \Omega_i(p)$, therefore consider integrating either one of these functions with a density of states that is varies smoothly near $\Omega_i(p)$,
\be \int_{-\infty}^\infty \rho(E_S) |F_i|^2 dE_S \approx \rho(\Omega_i) \int_{-\infty}^\infty |F_i|^2 dE_S \approx \rho(\Omega_i)~ 2\pi T_\pi(k) \Big[1-e^{-\Gamma_\pi(k)\,t_\mu}\Big] \label{intfun} \ee where we used eqn. (\ref{integF}) which implies the identification   (\ref{aprointF}). Similarly for large $t_D - t_\mu$
\be   \int_{-\infty}^\infty \rho(E_D) |G_i|^2 dE_D \approx \rho(\Omega_i) \int |G_i|^2 dE_D \approx \rho(\Omega_i)~ 2\pi (t_D-t_\mu)   \label{intGfun} \ee hence just as in Fermi's Golden rule we identify
\be |G_i|^2 = 2\pi \,(t_D-t_\mu) \,\delta(E_D-\Omega_i)\,. \label{FGR}\ee

Furthermore    for $\delta m^2 \lesssim 1\,(\mathrm{eV})^2$ which is  the putative mass range of sterile neutrinos to explain the short baseline anomalies, and typical neutrino energies $\oO \gtrsim 3\,\textrm{Mev}$ with the lower range applying to reactor neutrinos, then $\Delta \lesssim 10^{-7}\,(\mathrm{eV})$ and $\Delta/\oO \lesssim 10^{-14}$. In typical neutrino experiments, the energy spectrum is measured with a finite resolution and  ``binned'', namely integrated over an energy range determined by the resolution, however, such resolution is always much larger than $10^{-7}\, \mathrm{eV}$ and in all experiments the relative error in energy resolution $\Delta E/E \gg 10^{-14}$. The point is that
the measurement resolution is nowhere near enough to discriminate an energy difference $\Delta(p)$ between the energy eigenstates in the ``binning'', and replacing $\delta(E_{S,D}-\oO \pm \Delta) \rightarrow \delta(E_{S,D}-\oO )$ is an excellent approximation. Therefore we can safely replace

\be |F_1|^2 = |F_2|^2 = 2\pi T_\pi(k) \Big[1-e^{-\Gamma_\pi(k)\,t_\mu}\Big] \,\delta\big(E_S-\oO(p)\big) = 2\pi T_\pi(k) \Big[1-e^{-\Gamma_\pi(k)\,t_\mu}\Big] \,\delta\big(\mathcal{E}_S\big)\,. \label{aprox1}\ee

\be |G_1|^2 = |G_2|^2 = 2\pi (t_D-t_\mu) \,\delta\big(E_D-\oO(p)\big)=2\pi (t_D-t_\mu) \,\delta\big(\mathcal{E}_D\big)\,. \label{aprox2}\ee

The approximations above rely on that $F_{1,2};G_{1,2}$ are \emph{distributions} that are sharply peaked
and they must be understood as being integrated with density of states, which experimentally are insensitive to the energy difference $ \Delta(p)$ between the neutrinos. Under these approximations, consider the
 product
 \be F_2\,F^*_1 = \frac{1-2\,e^{i\Delta(p)t_\mu}\,e^{-\Gamma_\pi(k)t_\mu/2}\cos\big(\mathcal{E}_S\,t_\mu\big)+e^{2i\Delta(p)t_\mu}\,e^{-\Gamma_\pi(k)t_\mu}}{\Bigg(
\mathcal{E}_S-\Delta(p)-i\frac{\Gamma_\pi(k)}{2} \Bigg)\,\Bigg(
\mathcal{E}_S+\Delta(p)+i\frac{\Gamma_\pi(k)}{2} \Bigg)} \label{prodFs}\ee This function features two poles in the complex $\mathcal{E}_S$ plane at $\mathcal{E}_S= \pm (\Delta(p)+i \,\Gamma_\pi(k)/2)$, being integrated with a density of states that is insensitive to $\Delta(p)$ in the narrow width approximation
\be \int_{-\infty}^\infty \rho(\mathcal{E}_S) F_2 F^*_1\,d\mathcal{E}_S \simeq \rho(0) \int_{-\infty}^\infty F_2 F^*_1\,d\mathcal{E}_S \ee the integral can be done and we find
\be \int_{-\infty}^\infty F_2 F^*_1\,d\mathcal{E}_S  = 2\pi~ T_\pi(k)~\Bigg[ \frac{1 + i\frac{2\Delta(p)}{\Gamma_{\pi}(k)}}{1+ \frac{4\Delta^2(p)}{\Gamma^2_{\pi}(k)}}\Bigg]~ \Bigg[1- e^{2i\Delta(p)\,t_\mu}\,e^{-\Gamma_\pi(k)\,t_\mu} \Bigg] \label{Fprods}\ee using these results for $\Gamma_\pi =0$ we obtain \be \int_{-\infty}^\infty \rho(\mathcal{E}_D) G_2 G^*_1\,d\mathcal{E}_D \simeq \rho(0)  ~ (2\pi)~ e^{-i\Delta(p)(t_D+t_\mu)}~ \frac{\sin\big[\Delta(p)(t_D-t_\mu)\big]}{\Delta(p)}\,  \label{Gprods} \ee

Therefore, under the assumption that the experimental ``binning'' is insensitive to the neutrino energy difference we can safely approximate the direct terms as,
\bea |F_1|^2 & = &  |F_2|^2= 2\pi T_\pi(k) \Big[1-e^{-\Gamma_\pi(k)\,t_\mu}\Big] \,\delta\big(\mathcal{E}_S\big) \nonumber \\ |G_1|^2  & = &  |G_2|^2 =  2\pi (t_D-t_\mu) \,\delta\big(\mathcal{E}_D\big)\,, \label{aproxFGs}\eea and for the interference terms

\bea F_2 F^*_1 & = &  2\pi \,T_\pi(k)~\Bigg[ \frac{1 + i\mathcal{R}}{1+\mathcal{R}^2} \Bigg]~ \Bigg[1- e^{2i\Delta(p)\,t_\mu}\,e^{-\Gamma_\pi(k)\,t_\mu} \Bigg]\,\delta(\mathcal{E}_S) \nonumber \\ G_2 G^*_1 & = &  2\pi\, e^{-i\Delta(p)(t_D+t_\mu)}~ \frac{\sin\big[\Delta(p)(t_D-t_\mu)\big]}{\Delta(p)} \, \delta(\mathcal{E}_D) \label{aproxf1f2g1g2} \eea
where we have introduced the ratio
\be \mathcal{R} = \frac{2\,\Delta(p)}{\Gamma_\pi(k)} = \Bigg( \frac{\delta\,m^2}{2\,M_\pi\,\Gamma_0}\Bigg)\,\frac{E_{\pi}(k)}{\oO(p)} \,,\label{ratio}\ee where $\Gamma_0$ is the rest-frame decay width of the pion.

With these approximations, the number (density) of charged leptons ( muons ) measured at disentanglement
time $t_\mu$ (\ref{numnu}), is obtained by using the approximation (\ref{aprointF}) and $\Omega_i(p) \simeq E_\nu(p)$, we find
  \be \mathcal{N}_\mu(\vec{Q},t_\mu) = \frac{ 2\pi\,g^2_\pi \,  T_\pi(k) \,\Big[1-e^{-\Gamma_\pi(k)\,t_\mu}\Big] }{\Big[8VE_\pi(k)E_\mu(Q)\oO(p)\Big] } \,\delta\big(\mathcal{E}_S\big)\,, \label{numnuappx}\ee

  and     the number (density) of charged leptons measured at the detector   is given by
\be  \mathcal{N}_\mu(t_D)   =     \mathcal{N}_\mu(t_\mu)~d\Gamma_{\nu \rightarrow W\,\mu} \,\Bigg\{(\cos^4(\theta)+\sin^4(\theta))\,
(t-t_\mu)  +   \frac{1}{2}\, {\sin^2(2\theta)}\mathcal{T}[t_D,t_\mu] \Bigg\} \label{numutD}\ee
and
\be  \mathcal{N}_e(t_D)   =     \mathcal{N}_\mu(t_\mu) ~d\Gamma_{\nu \rightarrow W\,e} \, \frac{1}{2}\, {\sin^2(2\theta)}\, \Bigg\{(t-t_\mu)   -  \mathcal{T}[t_D,t_\mu] \Bigg\} \label{numetD}\ee
 where
 \bea \mathcal{T}[t_D,t_\mu]  & = &  \frac{1}{1+\mathcal{R}^2}~\frac{1}{1-e^{-\Gamma_\pi(k)t_\mu}}~
 \frac{\sin\big[\Delta(p)(t_D-t_\mu)\big]}{\Delta(p)}~
 \Bigg\{\Big(\cos[\Delta(p)(t_D+t_\mu)]+\mathcal{R}\,\sin[\Delta(p)(t_D+t_\mu)]\Big)\nonumber \\ & - & e^{-\Gamma_\pi(k)t_\mu}\Big(\cos[\Delta(p)(t_D-t_\mu)]+\mathcal{R}\,\sin[\Delta(p)(t_D-t_\mu)]\Big)
 \Bigg\}\label{capT}\eea and
 \be d\Gamma_{\nu \rightarrow W\,l}= \Bigg[ \frac{2\pi\,g^2\,\delta(E_D-\oO(p))}{8VE_W(k_w)E_l(q_l)\oO(p)}\Bigg]\,,\label{dG}\ee   is the differential charged lepton production rate from the reaction $\nu \rightarrow W\,l$ for a neutrino of energy $\oO$ at the detector.

 These expressions become more familiar if we calculate the detection \emph{rate}   as is the usual procedure in S-matrix theory, we find the simpler results
\be \frac{d\,\mathcal{N}_\mu(t_D)}{dt_D}= \mathcal{N}_\mu(t_\mu)\,d\Gamma_{\nu \rightarrow W\,\mu}\,\mathcal{P}_{\mu \rightarrow \mu}(t_D) \label{ratemu} \ee
\be \frac{d\,\mathcal{N}_e(t_D)}{dt_D}= \mathcal{N}_\mu(t_\mu)\,d\Gamma_{\nu \rightarrow W\,e}\,\mathcal{P}_{\mu \rightarrow e}(t_D) \,,\label{ratee} \ee    where the survival (disappearance) and appearance probabilities are
 \bea \mathcal{P}_{\mu \rightarrow \mu}(t_D) & = & 1-\frac{1}{2}\sin^2(2\theta) \Bigg\{1-\frac{1}{1+\mathcal{R}^2}\frac{1}{1-e^{-\Gamma_\pi(k)t_\mu}}\Bigg[ \Bigg(\cos[2\,\Delta(p)\,t_D]+\mathcal{R}\sin[2\,\Delta(p)\,t_D]\Bigg)-\nonumber \\ &&
e^{-\Gamma_\pi(k)t_\mu}\Bigg(\cos[2\,\Delta(p)\,(t_D-t_\mu)]+\mathcal{R}\sin[2\,\Delta(p)\,(t_D-t_\mu)]\Bigg) \Bigg]  \Bigg\} \label{Pmumuac}\eea
\bea \mathcal{P}_{\mu \rightarrow e}(t_D) & = &   \frac{1}{2}\sin^2(2\theta) \Bigg\{1-\frac{1}{1+\mathcal{R}^2}\frac{1}{1-e^{-\Gamma_\pi(k)t_\mu}}\Bigg[ \Bigg(\cos[2\,\Delta(p)\,t_D]+\mathcal{R}\sin[2\,\Delta(p)\,t_D]\Bigg)-\nonumber \\ &&
e^{-\Gamma_\pi(k)t_\mu}\Bigg(\cos[2\,\Delta(p)\,(t_D-t_\mu)]+\mathcal{R}\sin[2\,\Delta(p)\,(t_D-t_\mu)]\Bigg) \Bigg]  \Bigg\} \label{Pmueac}\eea

These expressions are in agreement with the previous discussion: when $\Gamma_\pi \gg 2\,\Delta(p)$ and $t_D \gg t_\mu$ (or $t_{osc} \gg t_\mu$)  the mass eigenstates cannot be discriminated during the lifetime of the source, $\mathcal{R} \rightarrow 0$  and
the appearance and disappearance probabilities are given by the  usual result
\be \mathcal{P}_{\mu \rightarrow \mu}(t_D) = 1- \sin^2(2\theta)\,\sin^2\Big[ \frac{\delta m^2}{4\oO(p)}\,t_D\Big]~~;~~\mathcal{P}_{\mu \rightarrow e}(t_D) =   \sin^2(2\theta)\,\sin^2\Big[ \frac{\delta m^2}{4\oO(p)}\,t_D\Big]\,. \label{Pfam}\ee However, in the opposite limit $2\,\Delta(p) \gg \Gamma_\pi(k)$,   namely $\mathcal{R}\gg 1$ or  $t_{osc} \simeq t_\mu$  the mass eigenstates are completely separated by the time evolution and the oscillation probabilities are suppressed.


The origin of the discrepancy between between the probabilities (\ref{Pmumuac},\ref{Pmueac}) and the
familiar results given by (\ref{Pfam}) is traced to the interference term $\mathrm{Re}[F_2F^*_1G_2G^*_1]$, the functions $F_j,G_j$ are completely determined by the time evolution of the quantum state and describe the approximate energy conservation at the production and detection vertices. The functions $F_{1,2}$ describe the initial correlations in the entangled quantum state (\ref{entstate}).

Therefore the physical reason behind the difference between the probabilities (\ref{Pmumuac},\ref{Pmueac}) and (\ref{Pfam}) is that the time scales associated with the lifetime of the source and the entanglement of the charged lepton define energy uncertainties which determine whether the mass eigenstates are separated during these time scales or not. Short lifetimes and disentanglement time scales ($T_\pi(k) \,;\,t_\mu \ll t_{osc} $) introduce large energy uncertainties and the mass eigenstates are ``blurred'' into a Pontecorvo state which yields the usual quantum mechanical result (\ref{Pfam}). In the opposite limit, long lifetime and disentanglement scales ($T_\pi(k) \,;\,t_\mu \gg t_{osc} $) lead to small energy uncertainties and the mass eigenstates are separated leading to decoherence which is manifest in the expressions (\ref{Pmumuac},\ref{Pmueac}) in terms of $\mathcal{R}$ and $t_\mu$.


It is important to highlight that the factorization in (\ref{ratemu},\ref{ratee}) is a direct consequence of the fact that the binning or energy resolution in all current experiments cannot distinguish the energy difference between the mass eigenstates $\Delta(p) \lesssim 10^{-7}\,\mathrm{eV}$ and energy resolutions $\Delta(p)/E_\nu(p) \lesssim 10^{-14}$ therefore the approximations (\ref{aprox1}-\ref{aprox2}) are amply justified.

The interplay between the lifetime of the source and the disentanglement time scale and the suppression of the oscillatory component of the transition probabilities is more clearly exhibited in two simple cases:

\vspace{2mm}

\textbf{Case I:   $\mathbf{\Gamma_\pi(k)\,t_\mu \gg 1}$:} This corresponds to a disentanglement time scale much larger than the lifetime of the source, in which case the energy uncertainty is determined by $\Gamma_\pi$.

 In this case
\be  \mathcal{P}_{\mu \rightarrow \mu}(t_D)   =   1-\frac{1}{2}\sin^2(2\theta) \Bigg\{1-\frac{1}{1+\mathcal{R}^2} \Bigg[ \cos[2\,\Delta(p)\,t_D]+\mathcal{R}\sin[2\,\Delta(p)\,t_D]  \Bigg]  \Bigg\} \label{PmumuacI}\ee
\be  \mathcal{P}_{\mu \rightarrow e}(t_D)   =     \frac{1}{2}\sin^2(2\theta) \Bigg\{1-\frac{1}{1+\mathcal{R}^2} \Bigg[ \Bigg(\cos[2\,\Delta(p)\,t_D]+\mathcal{R}\sin[2\,\Delta(p)\,t_D]\Bigg)  \Bigg]  \Bigg\} \label{PmueacII}\ee

This expressions make clear that when $\Delta(p) \gg \Gamma_\pi(k)$, namely for $\mathcal{R}\gg 1$ oscillations are suppressed in agreement with the analysis presented in the previous section. The suppression is a consequence of the separation of the mass eigenstates and the ensuing loss of coherence.

Note that whereas the usual expressions (\ref{Pfam}) valid for $\mathcal{R}=0$  are such that $\mathcal{P}_{\mu\rightarrow \mu} \rightarrow 1~ ; ~\mathcal{P}_{\mu\rightarrow e} \rightarrow 0$ as $t_{D} \rightarrow 0$, for $\mathcal{R} \neq 0$ this is not the case, because the expressions (\ref{PmumuacI},\ref{PmueacII}) only hold for $t_D>t_\mu \gg T_\pi$.

\vspace{2mm}

\textbf{Case II:   $\mathbf{\Gamma_\pi(k)\,t_\mu \ll 1}$:}

\vspace{2mm}

In this case the disentanglement time scale is much shorter than the lifetime of the source, namely the source is nearly stationary during the time scale of disentanglement and we can simply approximate this case by taking $\Gamma_\pi(k) \rightarrow 0$. This approximation correctly describes the fact that the main energy uncertainty is determined by $2\pi / t_\mu$ as explained following equations  (\ref{modf12small},\ref{modf22small}). In this case we find

\be F_j = e^{-iE_S\,t_\mu/2}\,e^{i\Omega_j\,t_\mu/2}\, 2i\, \frac{\sin\Big[(E_S-\Omega_j) t_\mu/2\Big]}{\Big[E_S-\Omega_j \Big]} \label{Fscase2} \ee and invoking the same approximations as in
the previous case we find
\be |F_1|^2=|F_2|^2 = 2\pi\,t_\mu\,\delta(E_S-\oO(p)) \label{case2ap}\ee
 \be \mathcal{N}_\mu(\vec{Q},t_\mu) = \frac{ 2\pi\,g^2_\pi \,  t_\mu \,\delta\big(E_S-\oO(p)\big)}{\Big[8VE_\pi(k)E_\mu(Q)\oO(p)\Big] } \,.\label{numnuappx2}\ee The interference term is given by
 \bea \mathrm{Re}
\Big[ F_2\,F^*_1 \,G_2\,G^*_1
\Big] &  = &  \cos(\Delta(p)\,t_D)\, \frac{\sin(\Delta(p)\,t_\mu)}{\Delta(p)}\,\frac{\sin(\Delta(p)\,(t_D-t_\mu))}{\Delta(p)}\nonumber \\& \times & (2\pi)\,\delta(E_S-\oO(p))
 \,(2\pi)\,\delta(E_D-\oO(p))\,,\label{realpart2}\eea leading to the final results
 \be \frac{d\,\mathcal{N}_\mu(t_D)}{dt_D}= \mathcal{N}_\mu(t_\mu)\,d\Gamma_{\nu \rightarrow W\,\mu}\,\mathcal{P}_{\mu \rightarrow \mu}(t_D) \label{ratemu2} \ee
\be \frac{d\,\mathcal{N}_e(t_D)}{dt_D}= \mathcal{N}_\mu(t_\mu)\,d\Gamma_{\nu \rightarrow W\,e}\,\mathcal{P}_{\mu \rightarrow e}(t_D)\,, \label{ratee2} \ee     where the survival (disappearance) and appearance probabilities are
\be \mathcal{P}_{\mu \rightarrow \mu}(t_D) = 1-\frac{1}{2}\sin^2(2\theta)\Bigg[1-\frac{\sin\big(\Delta(p)\,t_\mu\big)}{\Delta(p)\,t_\mu}\,\cos\Big[2\,\Delta(p)\big(t_D-\frac{t_\mu}{2}\big)\Big]  \Bigg] \label{Pmumu2}\ee
\be \mathcal{P}_{\mu \rightarrow e}(t_D) =  \frac{1}{2}\sin^2(2\theta)\Bigg[1-\frac{\sin\big(\Delta(p)\,t_\mu\big)}{\Delta(p)\,t_\mu}\,\cos\Big[2\,\Delta(p)\big(t_D-\frac{t_\mu}{2}\big)\Big]  \Bigg]\,. \label{Pmue2}\ee As $\Delta(p)\,t_\mu \rightarrow 0$, namely when $t_\mu/t_{osc}\rightarrow 0$, the expressions for the   probabilities become the familiar ones, but for $\Delta(p)\,t_\mu \gg 1$ the expressions above display   \emph{two} sources of suppression through entanglement: the prefactor ${\sin\big(\Delta(p)\,t_\mu\big)}/{\Delta(p)\,t_\mu}$, and also a shortening of the effective baseline from
$L = c\,t_D$ to $L_{eff}= c\,(t_D-\frac{t_\mu}{2})$  .



\section{Implications for accelerator and reactor experiments:}\label{expts}
The discussion of the previous sections hinges on two generation mixing, however, if sterile neutrinos
are the correct explanation of the short-baseline anomalies then
\be \sin^2(2\theta) \rightarrow 4 |U_{e 4}|^2\,|U_{\mu 4}|^2 \,.\label{changeang}\ee

It is convenient to write the probabilities in terms of the baseline $L = c\,t_D$ and introduce the disentanglement length of the muon $L_d= c\,t_\mu$,  writing as usual
\be \Delta(p)\,t_D = 1.27~ \frac{\delta m^2}{\mathrm{eV}^2}\,\frac{L/\mathrm{m}}{E_\nu(p)/\mathrm{MeV}}~~;~~\Delta(p)\,t_\mu = 1.27\, \frac{\delta m^2}{\mathrm{eV}^2}\,\frac{L_d/\mathrm{m}}{E_\nu(p)/\mathrm{MeV}}\label{defi}\ee and
\be \Delta(p)=1.27\, \frac{\delta m^2}{\mathrm{eV}^2}\,\frac{\mathrm{MeV}}{E_\nu(p)} \equiv \frac{\pi}{L_{osc}/\mathrm{m}} \,. \label{Losc}\ee The disappearance and appearance probabilities are given by
 \bea \mathcal{P}_{\mu \rightarrow \mu}(L) & = & 1-\frac{1}{2}\sin^2(2\theta) \Bigg\{1-\frac{1}{1+\mathcal{R}^2}\frac{1}{1-e^{- L_d/L_\pi(k)}}\Bigg[ \Bigg(\cos[2\,\Delta(p)\,L]+\mathcal{R}\sin[2\,\Delta(p)\,L]\Bigg)-\nonumber \\ &&
e^{-L_d/L_\pi(k)}\Bigg(\cos[2\,\Delta(p)\,(L-L_d)]+\mathcal{R}\sin[2\,\Delta(p)\,(L-L_d)]\Bigg) \Bigg]  \Bigg\} \label{PmumuacL}\eea
\bea \mathcal{P}_{\mu \rightarrow e}(L) & = &   \frac{1}{2}\sin^2(2\theta) \Bigg\{1-\frac{1}{1+\mathcal{R}^2}\frac{1}{1-e^{-L_d/L_\pi(k)}}\Bigg[ \Bigg(\cos[2\,\Delta(p)\,L]+\mathcal{R}\sin[2\,\Delta(p)\,L]\Bigg)-\nonumber \\ &&
e^{- L_d/L_\pi(k)}\Bigg(\cos[2\,\Delta(p)\,(L-L_d)]+\mathcal{R}\sin[2\,\Delta(p)\,(L-L_d)]\Bigg) \Bigg]  \Bigg\} \label{PmueacL}\eea where $L_\pi(k)=c/\Gamma_\pi(k)$  is the decay length\footnote{For $E_\pi(k)  \gtrsim 1\,\mathrm{GeV}$ the Lorentz factor $\gamma \gtrsim 7$ and we approximate $\beta \sim 1$.} of the parent particle (pion).

For neutrinos produced from pion decay the ratio $\mathcal{R}$  (\ref{ratio}) becomes
\be \mathcal{R}= 2\pi ~ \frac{L_\pi(k)}{L_{osc}(E_\nu)} = 0.14~~\frac{\delta m^2}{ (\mathrm{eV})^2}~\Bigg(\frac{E_\pi(k)}{E_\nu(p)} \Bigg) \label{ratiopion}\ee  therefore,  $\mathcal{R} \simeq 1$ for $\delta m^2 \simeq (\mathrm{eV})^2 $ which is the range of masses for sterile neutrinos that \emph{could} solve the short-baseline anomalies, and in the case of  MiniBooNE  $ 1 \lesssim E_\pi/E_\nu \lesssim 6$.


Remarkably, eqn. (\ref{PmumuacL}) is exactly the same as eqn. (23) in ref.\cite{smirnov2} where $\mathcal{R}$ is equivalent to the quantity $\xi$ and $L_\mu$ replaces the ``pipeline'' $l_d$ in this reference. Hence, the result of ref.\cite{smirnov2} can be interpreted as disentangling the muon at the distance $l_d$ which is identified with the ``pipeline''.


Both LSND and MiniBooNE are designed with $(L/m)/(E_\nu/(\mathrm{MeV}) \simeq 1$.

At MiniBooNE a neutrino beam is obtained from   pions that decay in a decay ``pipe'' $\simeq 50 \, \mathrm{m}$ long,   and neutrinos go through $\simeq 500 \, \mathrm{m}$ of ``dirt'' before reaching the
detector determining a baseline   $L\,\sim 550 \, \mathrm{m}$ with a peak energy in the neutrino spectrum at about $600-1000\,\mathrm{MeV}$. At MiniBooNE muons with $\sim \mathrm{GeV}$ energy are stopped at a distance
$\simeq 4\,\mathrm{m}$ in the ``dirt'' thus $L_d \simeq 54\,\mathrm{m}$\footnote{The author is indebted to William C. Louis III for extensive correspondence   clarifying these experimental aspects of the MiniBooNE experiment.}.

\begin{figure}[h!]
\begin{center}
\includegraphics[height=3in,width=3in,keepaspectratio=true]{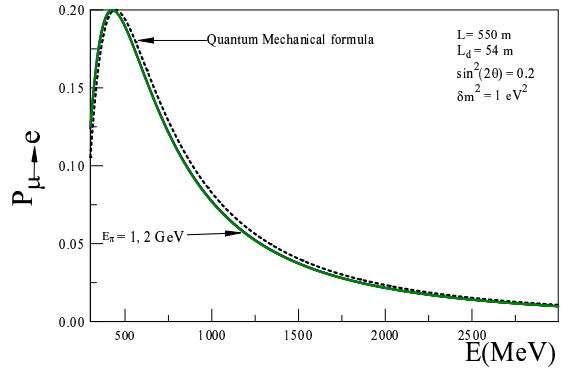}
\caption{Appearance $\mathcal{P}_{\mu \rightarrow e}$ probability vs. $E_\nu\,(\mathrm{MeV})$ for MiniBooNE parameters. The solid line(s) correspond to (\ref{PmueacL}) for $E_\pi(k) = 1,2 \,\mathrm{GeV}$ (indistinguishable on the scale of the figure). The dashed line corresponds to the quantum mechanical probability $\mathcal{P}_{\mu \rightarrow e} $ (\ref{Pfam}).  } \label{fig:miniboonerev}
\end{center}
\end{figure}

Fig. (\ref{fig:miniboonerev}) displays these probabilities for the set of parameters consistent with (one) sterile neutrino with $\delta m^2 \simeq 1\,\mathrm{eV}^2~;~\sin^2(2\theta) = 0.2$ and MiniBooNE baseline and range of neutrino energies.

\begin{figure}[h!]
\begin{center}
\includegraphics[height=3in,width=3in,keepaspectratio=true]{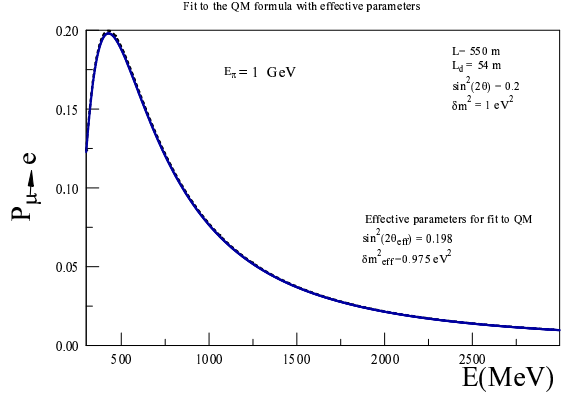}
\caption{ Fit of  $\mathcal{P}_{\mu \rightarrow e}$  vs. $E_\nu\,(\mathrm{MeV})$ for MiniBooNE parameters and $E_\pi= 1\,\mathrm{GeV}$. For $\sin^2(2\theta) =0.2\,,\,\delta m^2 = 1\,\mathrm{eV}^2$ the fit yields $\sin^2(2\theta_{eff}) =0.198\,,\,\delta m^2_{eff} = 0.975\,\mathrm{eV}^2$ } \label{fig:revminiboonefit}
\end{center}
\end{figure}

The  figure shows that the appearance probability is suppressed as compared to the quantum mechanical result, the suppression being more pronounced at smaller energy where $\mathcal{R}$ is larger (see below). Although for $\mathcal{R} \neq 0$ the probabilities cannot be fit by the usual quantum mechanical result in the full energy range, a fit of the form
\be \mathcal{P}_{\mu \rightarrow e}(L) =   \sin^2(2\theta_{eff})\,\sin^2\Bigg[ \frac{\delta m^2_{eff}}{4\oO}\,L\Bigg] \label{Pfameff}\ee in a restricted energy range
would lead to
\be \sin^2(2\theta_{eff}) < \sin^2(2\theta)~~;~~\delta m^2_{eff} < \delta m^2 \label{effpars}\ee
       as can be seen from the position of the maxima of the appearance probability: lower in amplitude (smaller mixing angle) and moving towards smaller energy (smaller $\delta m^2$). For the case of MiniBooNE the fit is shown in fig.\ref{fig:revminiboonefit}

Because in this situation  $L_d \simeq L_\pi(k) \ll L$     decoherence from the source lifetime or entanglement does not lead to  experimentally substantial corrections.

Although not relevant for the MiniBooNE experiment, but as an illustrative example  to display the effects of decoherence on the transition probabilities as a consequence of long distance entanglement, we consider the case $L_d \gg L_\pi(k)$, in which case the probabilities (\ref{PmumuacL},\ref{PmueacL}) simplify to

 \be  \mathcal{P}_{\mu \rightarrow \mu}(L)   =   1-\frac{1}{2}\sin^2(2\theta) \Bigg\{1-\frac{1}{1+\mathcal{R}^2} \Bigg[ \Bigg(\cos[2\,\Delta(p)\,L]+\mathcal{R}\sin[2\,\Delta(p)\,L]\Bigg) \Bigg] \Bigg\} \label{Pmumulong}\ee
\be  \mathcal{P}_{\mu \rightarrow e}(L)   =     \frac{1}{2}\sin^2(2\theta) \Bigg\{1-\frac{1}{1+\mathcal{R}^2} \Bigg[ \Bigg(\cos[2\,\Delta(p)\,L]+\mathcal{R}\sin[2\,\Delta(p)\,L]\Bigg) \Bigg\} \label{Pmuelong}\ee

\begin{figure}[h!]
\begin{center}
\includegraphics[height=3in,width=3in,keepaspectratio=true]{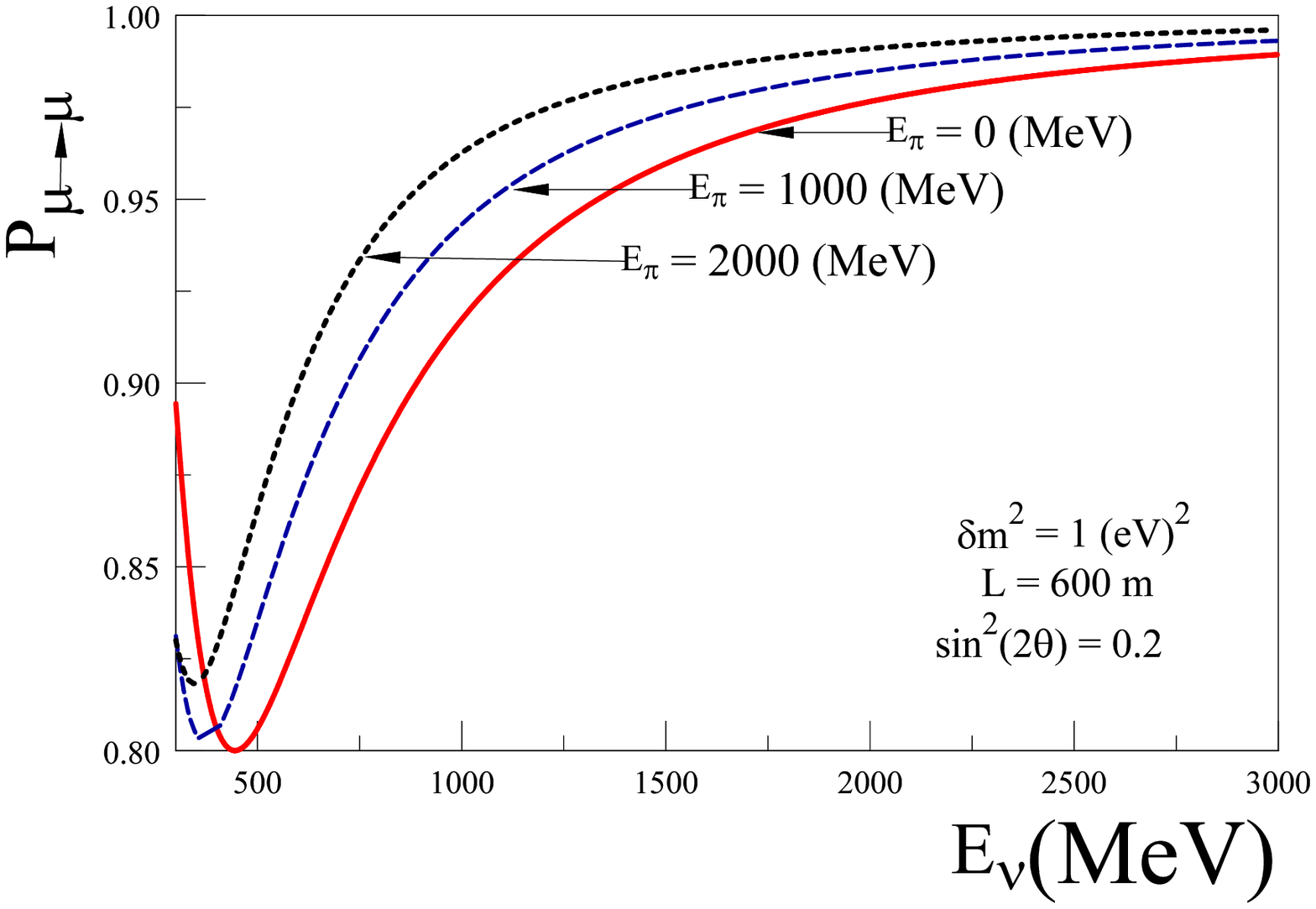}
\includegraphics[height=3in,width=3in,keepaspectratio=true]{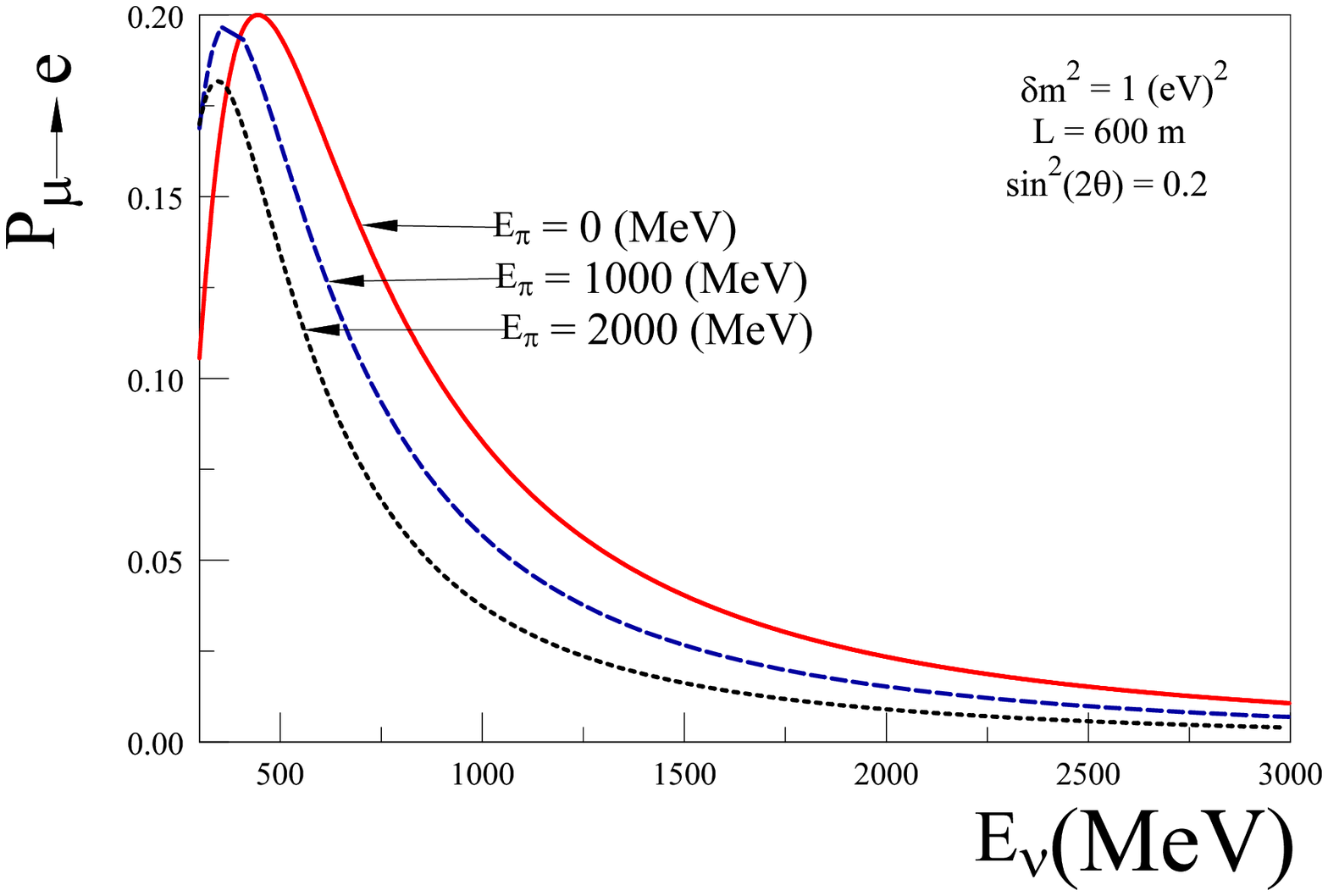}
\caption{Disappearance $\mathcal{P}_{\mu \rightarrow \mu}$ and appearance $\mathcal{P}_{\mu \rightarrow e}$ probabilities vs. $E_\nu\,(\mathrm{MeV})$ for $L=600\,\mathrm{m}$\,,\,$\delta m^2 = 1\,\mathrm{eV}^2$\,,\,$\sin^2(2\theta)=0.2$. The value $E_\pi =0$ refers to $\mathcal{R}=0$,   the usual quantum mechanical result for the the probabilities.  } \label{fig:miniboone}
\end{center}
\end{figure}

These are displayed in fig. (\ref{fig:miniboone}) for $L=600\,\mathrm{m}\,,\,\delta m^2 = 1\,\mathrm{eV}^2\,,\,\sin^2(2\theta)=0.2\,,\,E_{\pi}(k)= 1,2 \,\mathrm{GeV}$.

\begin{figure}[h!]
\begin{center}
\includegraphics[height=3in,width=3in,keepaspectratio=true]{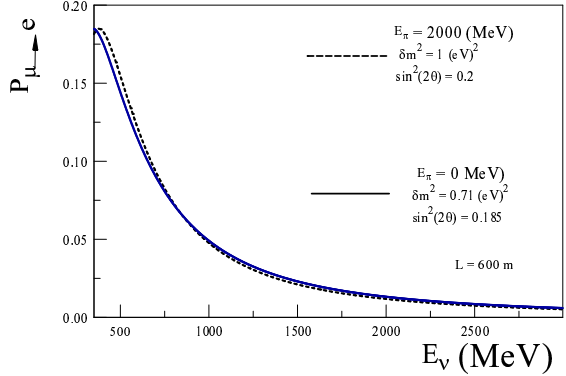}
 \caption{Disappearance $\mathcal{P}_{\mu \rightarrow \mu}$ for MiniBooNE parameters: the dashed line is the result with $E_\pi = 2000\,\mathrm{MeV}~;~\delta m^2= 1\,\mathrm{eV}^2~;~\sin^2(2\theta)=0.2$, the solid line is a fit with the  quantum mechanical probability (\ref{Pfameff}) with $\delta m^2_{eff} = 0.71\,\mathrm{eV}^2~;~\sin^2(2\theta_{eff})=0.185$. $L=600\,\mathrm{m}$.  } \label{fig:miniboonefit}
\end{center}
\end{figure}

 Figure  (\ref{fig:miniboonefit}) displays the appearance probability given by (\ref{Pmueac}), for the  parameters $E_\pi = 2000 \,\mathrm{MeV}~;~\delta m^2  =  1 \,\mathrm{eV}^2~;~L=600\,\mathrm{m}~~;~~\sin^2(2\theta)=0.2$ (solid line) and the best fit to the quantum mechanical probability (\ref{Pfameff}) resulting in $\delta m^2_{eff} = 0.71 \,\mathrm{eV}^2~~;~~\sin^2(2\theta_{eff}) = 0.185$. Several aspects are clarified by this example: i) the suppression by lifetime and disentanglement effects leads to an underestimate of both $\delta m^2,\sin^2(2\theta)$, ii) the fit is reliable only within an intermediate energy range, much less reliable in the low energy region, iii) the ratio $\mathcal{R}$ implies that there are more parameters than the amplitude $\sin^2(2\theta)$ and the ratio $L(m)/E_\nu(MeV)$.

Therefore, since the experimental data is always fit with the usual quantum mechanical formula,     the values of $\sin^2(2\theta)~;~\delta m^2$ from the fit actually correspond  to $\sin^2(2\theta_{eff})~;~\delta m^2_{eff}$ the above analysis leads to conclude that decoherence from the decay of the parent particle and the disentanglement of the charged lepton imply a \emph{larger value of the mixing angle and } $\delta m^2$ from those extracted from the fit to the usual quantum mechanical probability.

As shown above  for the parameters of MiniBooNE, decoherence through lifetime and entanglement effects yield
very small corrections, however the principal and fundamental observation remains, namely lifetime or disentanglement time scales similar to or larger than the oscillation time scale lead to decoherence and suppression of the appearance probabilities. A quantum mechanical fit yield effective values $\theta_{eff}\,,\,\delta m^2_{eff}$ which are \emph{smaller} than the actual values.

In our analysis we have assumed that the entangled quantum state arises from the two body decay of a parent particle, (here considered to be the pion), however at LSND the (anti) neutrino beam is produced by the \emph{three body} decay of a muon at rest, whereas at reactors  the (anti) neutrinos are produced via the   $\beta$ decay of long-lived unstable nuclei ${}^{235}\mathrm{U} \,, {}^{238}\mathrm{U} \,,{}^{239}\mathrm{Pu} \,, {}^{241}\mathrm{Pu}$\cite{flux1,reactor}. Although the actual calculation presented in the previous section for the exact entangled state does not directly apply to the description of the quantum states of neutrinos produced at LSND and of reactor experiments,   in absence of a more detailed understanding of the entangled quantum state resulting from the three body nuclear $\beta$ decay,  we will use the result (\ref{entstate}) with the caveat of possible corrections arising from three body phase space effects.

 At LSND muon antineutrinos are produced from $\pi^+ \rightarrow \mu^+\,\nu_\mu$ followed by $\mu^+ \rightarrow e^+ \nu_e \overline{\nu}_\mu$ where most of the muons decay at rest. The resulting $\overline{\nu}_\mu$ beam attains the maximum energy at the Michel end point $52.8\,\mathrm{MeV}$ and the liquid scintillator detector is located about $L = 30 \, \mathrm{m}$ from the neutrino source. Since $L \ll c\tau_\mu \simeq 660\,\mathrm{m}$ for LSND it follows that $\Gamma_\mu t_\mu \ll 1$ (the relevant decay width now is the muon's as the parent particle) and this corresponds to case II (nearly stationary case) with $L_d$ the disentanglement length. The same limit applies to reactor experiments where neutrinos are produced from nuclear $\beta$ decay of long-lived radiaoactive nuclei, therefore for LSND and reactor experiments the disappearance and appearance probabilities are given by,
 \be \mathcal{P}_{\mu \rightarrow \mu}(L) = 1-\frac{1}{2}\sin^2(2\theta)\Bigg[1-\frac{\sin\big(\Delta(p)\,L_d\big)}{\Delta(p)\,L_d}\,\cos\Big[2\,\Delta(p)\,L\,\Big(1-\frac{L_d}{2\,L}\Big)\Big]  \Bigg] \label{Pmumu2re}\ee
\be \mathcal{P}_{\mu \rightarrow e}(L) =  \frac{1}{2}\sin^2(2\theta)\Bigg[1-\frac{\sin\big(\Delta(p)\,L_d\big)}{\Delta(p)\,L_d}\,\cos\Big[2\,\Delta(p)\,L\,\Big(1-\frac{L_d}{2\,L}\Big)\Big]  \Bigg]\,. \label{Pmue2re}\ee
In ref.\cite{stock} it was also recognized that the muon lifetime does not affect the transition probabilities at LSND, however, the effect of disentanglement has not been previously recognized.

  In LSND, the detector  is at $L=30 \,\mathrm{m}$ from the neutrino source and is shielded by the equivalent of $9\,\mathrm{m}$ of steel\cite{lsnddet} which then should be taken as a figure of merit for $L_d \lesssim 20\,\mathrm{m}$.  At reactor experiments a figure of merit could be the size of the reactor core, at CHOOZ\cite{chooz} it is approximately $\simeq 4 \,\mathrm{m}$ with a baseline $100 \,\mathrm{m} \lesssim L \lesssim 1\,\mathrm{km}$, although, quite likely these figures of merit for $L_d$ overestimate  the disentanglement length scale both in LSND and in reactor experiments.  Unlike the case of MinibooNE where the suppression factor is mainly determined by the pion decay length, at LSND and reactor experiments the disentanglement scale $L_d$ is less certain.

\begin{figure}[h!]
\begin{center}
\includegraphics[height=3in,width=3in,keepaspectratio=true]{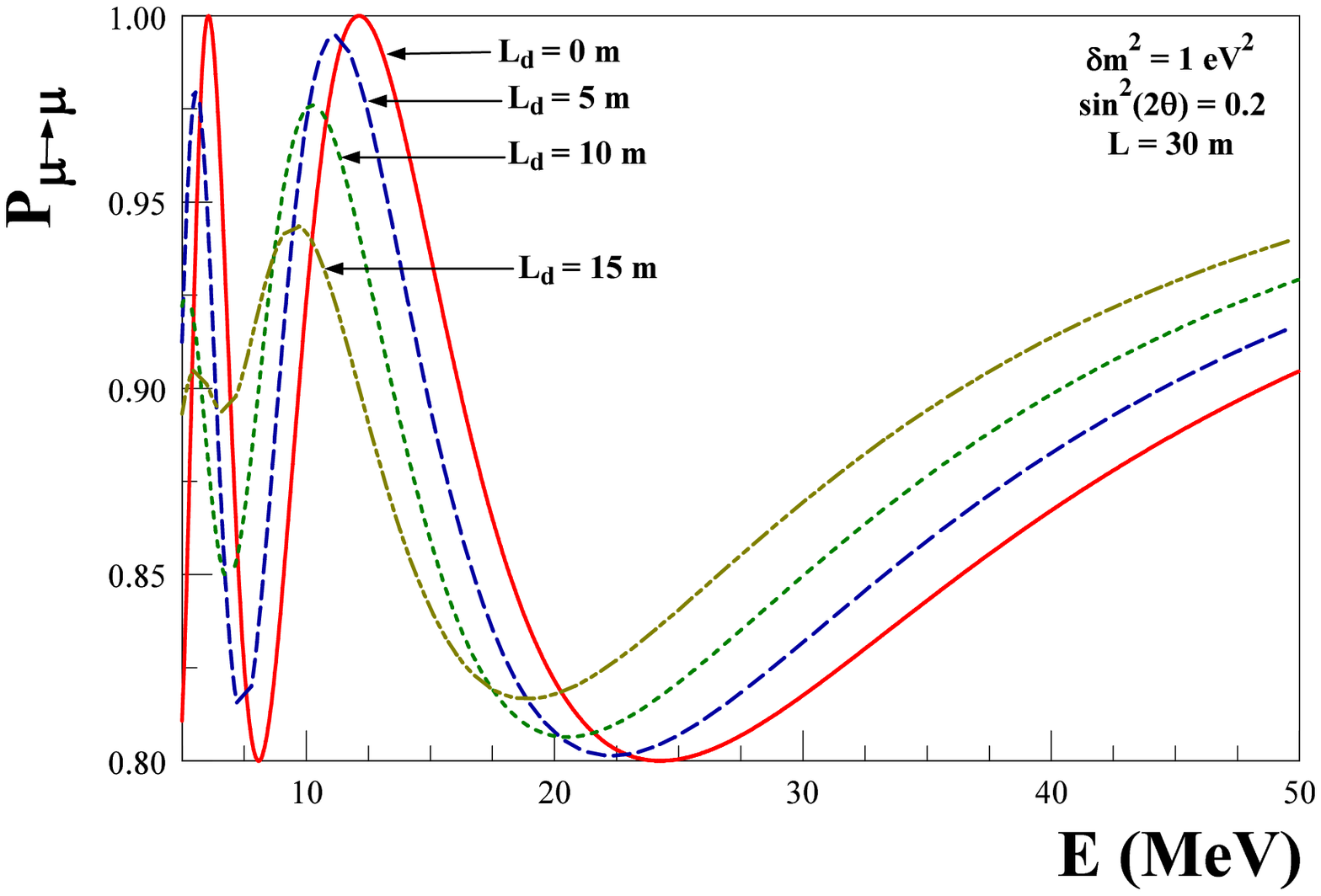}
\includegraphics[height=3in,width=3in,keepaspectratio=true]{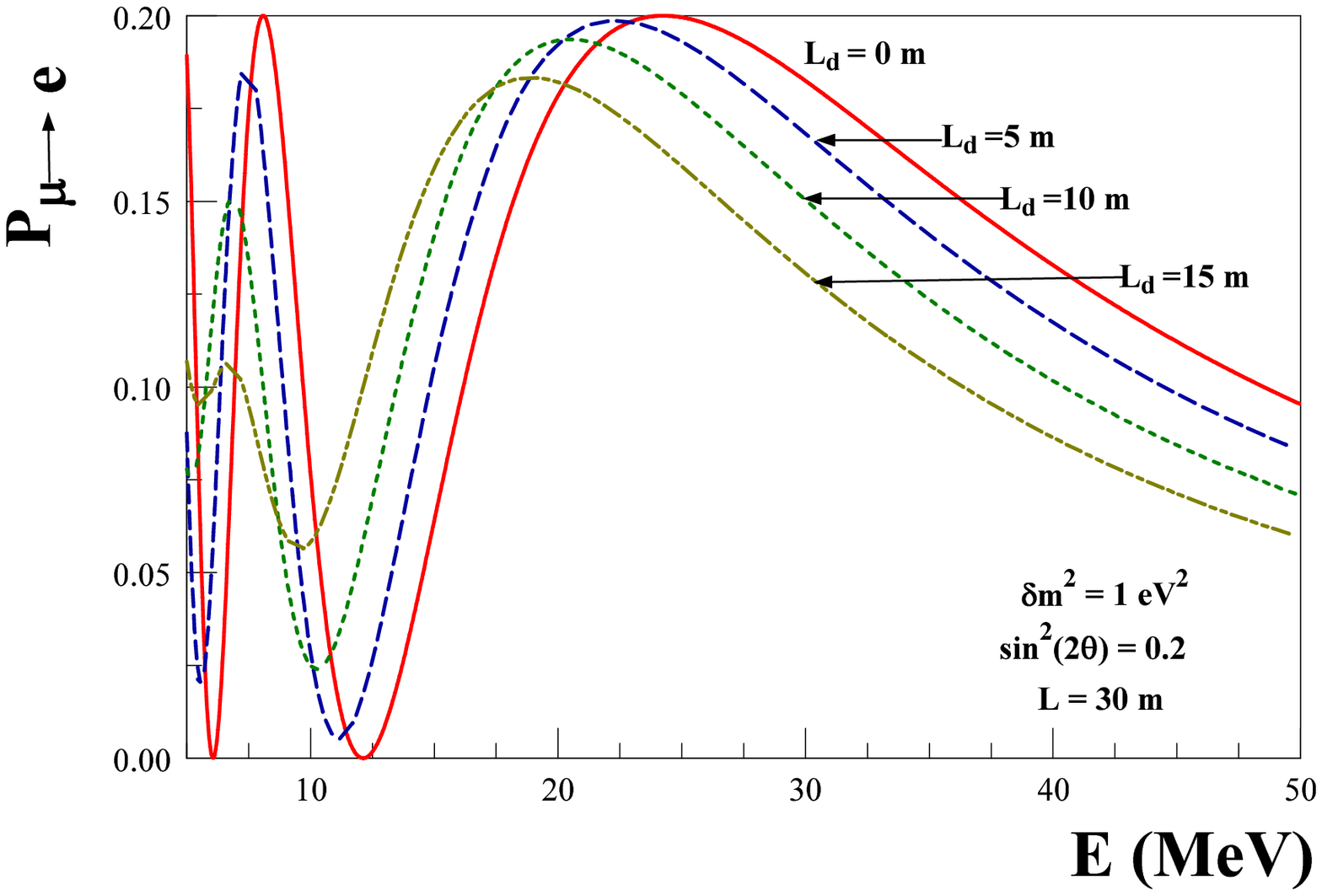}
\caption{Disappearance $\mathcal{P}_{\mu \rightarrow \mu}$ and appearance $\mathcal{P}_{\mu \rightarrow e}$ probabilities vs. $E_\nu\,(\mathrm{MeV})$ for LSND and reactor parameters. The value $L_d =0$ refers to the usual quantum mechanical result for the the probabilities.  } \label{fig:lsnd}
\end{center}
\end{figure}

 Thus we take $L_d$ as a parameter  and study the disappearance and appearance probabilities within the range $0\,\mathrm{m}  \lesssim L_d \lesssim 15\,\mathrm{m}$  to illustrate the consequences of decoherence from entanglement and to extract the main conclusions. These are displayed in fig. (\ref{fig:lsnd}).

These figures reveal a situation very similar to that analyzed above for MiniBooNE. Larger disentanglement lengths $L_d$ lead to a larger suppression of the appearance probability.  Similarly, a fit to the experimental data with the usual quantum mechanical appearance probability results in an \emph{underestimate} of both $\sin^2(2\theta)$ and $\delta m^2$ for the same reasons analyzed above.

\vspace{3mm}

\section{Comments on wave packets:}\label{wavepack}

\vspace{3mm}

The study in this article was restricted to plane waves to exhibit the main results and conclusions in the clearest possible setting. As has been argued in the literature\cite{kayser,rich,giunticohe,dolgov,beuthe,akmerev} wave packet localization is an important ingredient in the description of neutrino oscillations. The localization length both of the production and detection regions define momentum uncertainties that are important in the conceptual understanding of the interference phenomena.

Furthermore, in our calculation the disentanglement $t_\mu$ and detection $t_D$ times are sharp, this is a consequence of calculating the transition matrix elements in finite time intervals, however, the wave packet treatment smears these times over the time scale during which the wave packet overlaps with the detectors which is the appropriate physical description of the detection events.

The  analysis in\cite{kayser,rich,giunticohe,dolgov,beuthe,akmerev,boyho} (typically with Gaussian wave packets) clarifies that neutrino wave-packets evolve   semiclassically, the center moves as the front of a plane wave with the group velocity and is modulated by a Gaussian envelope which spreads through dispersion. Wave packets associated with the different mass eigenstates separate as they evolve with slightly different group velocities and when their separation becomes of the order of or larger than the width of the wave packet the overlap vanishes and oscillations are suppressed, typically exponentially in  the ratio $L^2/L^2_{coh}$ where $L_{coh} \simeq \sigma\,E^2_\nu/\delta m^2$ and $\sigma$ is the spatial localization scale of the wave packet. As discussed in \cite{bookchap8} the wave packet description also features another source of
decoherence in the \emph{localization term}, which suppresses coherence when    $\sigma > L_{osc}$.

However, it should be clear from the discussion and results presented above, that energy uncertainties from the width of the parent particle, disentanglement time scales, finite time intervals between production and detection and experimental measurements are sufficient to guarantee interference and oscillations. Entanglement over long distances and time scales introduces decoherence in a quantifiable manner. Introducing wave packets will modify the results only quantitatively but by no means fundamentally: a wave packet is a linear superposition of plane waves and the analysis for each plane wave described above can be generalized to such superposition. One  aspect that relies on a wave-packet description is the detection: the total number of events is obtained by the event \emph{rates} multiplied by the total time that the wave packet takes to pass through the detector. For ultrarelativistic neutrinos this is of order $\sigma/c$ since spreading through dispersion can be neglected on short baselines, therefore the total number of events is given by the rates (\ref{ratemu2},\ref{ratee2}) multiplied by $\sigma/c$, obviously this will not change the distortion of the spectrum determined by the oscillations in the appearance and disappearance probabilities. Another correction is the geometric flux factor which again for short baselines can be neglected. As found in refs.\cite{nauenberg,patkos,stock} including the lifetime of the source in the wave-packet evolution introduces another length scale (the decay length of the parent particle) which competes with the localization length of the wave-packet. As discussed above, wave packet localization will not  affect   oscillations unless the wave-packets corresponding to the different mass eigenstates begin
to separate. For $\delta m^2 \simeq 1\,\mathrm{eV}^2~;~E_\nu \sim \mathrm{MeV}$ and $L \sim 100\,\mathrm{m}$ the criterion for separation over the baseline would require a localization length $\sigma \lesssim 1 \,{\AA} $,  this estimate  is much larger than the nuclear radius  for unstable nuclei, thus decoherence via the separation of the wave-packets of mass eigenstates \emph{may} be another source of decoherence \emph{if} the localization length scale of the wave packets is of nuclear dimensions.

Thus we conclude that the results obtained with the plane wave analysis will apply \emph{vis a vis} to the case of wave packets, unless the source of decoherence associated with the separation of wave packets of mass eigenstates introduces enough decoherence as to dwarf the effects discussed here. On the short baseline experiments considered here this would require localization lengths $\lesssim 10^{-10} \,\mathrm{m}$ for reactor experiments and $\lesssim 10^{-15} \,\mathrm{m}$  for accelerator experiments.

Strengthening  these arguments require (and warrant) a full study of the  complete description of disentanglement and lifetime effects in a wave packet formulation. Of particular importance is whether for $\delta m^2 \sim \mathrm{eV}^2 ~;~E_\nu \sim \,\mathrm{few}\,\mathrm{MeV}$ wave packet localization on nuclear scales can be a source of  decoherence in reactor experiments. The results of this study will be presented elsewhere\cite{junme}.

\vspace{2mm}

\textbf{Wave packets vs. disentanglement:}

Decoherence through lifetime and disentanglement is fundamentally and conceptually different from decoherence in the wave packet formulation. Neutrino wavepackets manifestly describe single particle states that are \emph{spatially localized}, the spatial localization introduces uncertainty in the momentum, and in this formulation decoherence is a consequence of the separation \emph{in space} of the wave packets   associated with the different mass eigenstates. As explained in ref.\cite{bookchap8} there are two sources of decoherence: one resulting from the separation of the wave packets of different mass eigenstates through their different group velocity, and another determined by a \emph{localization term} (see eqn. (8.114) in ref.\cite{book5}) which results in decoherence for $\sigma \gg L_{osc}$.

Entanglement, on the other hand,  refers to the fact that the quantum state that results from the decay of the parent particle is a \emph{correlated many particle state}, the correlation between the charged lepton and the neutrino(s) is manifest in the coefficients $\mathcal{C}_i(\vk,\vq,\vp,t)$ in the quantum state (\ref{entstate}). These coefficients are time dependent and describe the \emph{approximate conservation of energy } at the production vertex.  A single particle neutrino state is obtained by projection of the charged lepton state, this projection is the quantum mechanical manifestation of the observation, absorption or decay of the charge lepton and \emph{disentangles} the (two body) quantum state at a time scale $t_\mu$.
  These correlations are precisely the origin of the terms $F_{1,2}$ which enter in the interference term  (\ref{interf}) and are, therefore, the origin of the difference with the familiar quantum mechanical result.
 In this description the lifetime of the source and $t_\mu$ determine \emph{energy uncertainties} as explained in the previous sections. Decoherence ensues when the energy uncertainty is much smaller than the energy separation between the mass eigenstates. This source of decoherence is obviously independent of the spatial localization of the
quantum state and is present even for plane waves, unlike wave packet decoherence.
Although decoherence in the wave packet and disentanglement formulations are physically and conceptually different, they are indeed complementary and \emph{both} will be present in a complete wave packet description of neutrino oscillations. For example as discussed in ref.\cite{bookchap8} \emph{if} a neutrino wavepacket produced by the decay of a parent particle of width $\Gamma$ is assigned a localization length $1/\Gamma$ then the condition for decoherence from the \emph{localization term} $\sigma \simeq L_{osc}$ becomes equivalent to $\Gamma \simeq \Delta(p)$ which is the condition which results from the disentanglement analysis in the case when the lifetime is \emph{shorter} than the disentanglement time scale. However, obviously this cannot be the case for reactor neutrinos since the lifetime of the parent particle is thousands of years and the relevant scale is
the disentanglement length scale as discussed above.

\section{Conclusions and further questions:}

Accumulating   evidence for anomalies in short-baseline experiments pointing towards a change in the current paradigm of neutrino oscillations resulting from the  mixing among three active species, will likely motivate further accelerator and reactor short baseline experiments. The firm assessment of   new ``sterile'' neutrinos as possible explanations of the data warrant a deeper understanding of  quantum coherence that determine the appearance and disappearance probabilities.

The realization that the   neutrino states produced in charged current interaction vertices are quantum entangled states of the neutrino and its flavor charged lepton partner call for a re-examination of the usual quantum mechanical description of neutrino oscillations as simple two level systems (for two neutrinos mixing). The measurement, absorption or decay of the charged lepton leads to the disentanglement of the quantum state, but the resulting neutrino state features the correlations from the prior entanglement.

The disentanglement of this correlated quantum state is a necessary condition for coherence between the mass eigenstates leading to oscillations, entanglement over long time scales project out energy eigenstates preventing oscillations. The usual ``Pontecorvo'' (quantum mechanical states) emerge if the disentanglement time scale is much smaller than the oscillation scale. This is a consequence of the time-energy uncertainty: for disentanglement time scales   shorter than the oscillation time, the uncertainty in energy cannot discriminate between the different mass eigenstates, the longer the entanglement time scale the smaller the energy uncertainty and the mass eigenstates become sharply defined in the correlated state leading to a suppression of the oscillation probability.

In this article we find that both the entanglement with the charged lepton and the lifetime of the source that produces the neutrino beam lead to a suppression of the appearance   probabilities. The relevant dimensionless parameter that quantifies decoherence by both effects is the ratio $\pi L_s/L_{osc}$ where $L_s$ is the \emph{smaller} between the decay length of the parent particle (source) and the disentanglement length scale.

We obtain the corrections to the disappearance and appearance probabilities both from entanglement and lifetime effects in a model which captures in a clear and reliably manner the main features of the production, evolution and detection of mixed states.

For MiniBooNE, the most important source of suppression is the decay length of the pions that produce the neutrino beam which is of the same order as the disentanglement length for the muons,  whereas at LSND and reactor experiments, the disentanglement distance is the relevant scale that determines the suppression, for LSND this is because neutrinos are produced by muons decaying at rest while in reactor experiments neutrinos are produced via $\beta$ decay of long lived radioactive sources, in both cases the disentanglement time scale is shorter than the lifetime of the source.

Short baseline experiments imply small $L_{osc}$ therefore the impact of disentanglement and source lifetime is larger in  these experiments. The  suppressions of the oscillation probabilities are more pronounced at lower energies and  are more dramatic for $\delta m^2 \sim 1\,\mathrm{eV}^2$ which is the mass range   for ``sterile'' neutrinos proposed as possible explanations of the short-baseline anomalies.

Our main results are the general disappearance and appearance probabilities   given by eqns. (\ref{PmumuacL},\ref{PmueacL}). These simplify to equations (\ref{Pmumu2re},\ref{Pmue2re})   when the disentanglement time scale is much shorter than the lifetime of the source, this is the case in reactor experiments (neutrinos at reactors are produced by $\beta$ decay of long lived radioactive nuclei) and at LSND. The determination of the scale $L_d$ is cleaner in accelerator experiments where the neutrino beam is produced by pion decay (either at rest or in flight). However, for MiniBooNE the corrections are relatively small because the disentanglement length scale is of the order of the pion decay length and both are much smaller than the baseline. In reactor experiments $L_d$ is more difficult to establish, a figure of merit is the size of the reactor core, but this estimate is probably too simplistic and overestimates the disentanglement length.

While the experimental impact of the corrections in current experiments is relatively small, this work suggests that in the analysis of the data, the issue of disentanglement length scale must be addressed for a consistent interpretation of the results. An important corollary of our results is  that fitting the experimental data with the usual quantum mechanical expressions for appearance and disappearance probabilities \emph{underestimates} both $\sin^2(2\theta)$ and $\delta m^2$, furthermore this fit to the data differs substantially at low neutrino energy from the correct expression for the probabilities that include both the lifetime and disentanglement suppression, since the suppression is larger at smaller energies (shorter $L_{osc}$).

An   aspect that remains to be explored further is the description of neutrino propagation in terms of wave packets: the source and detector are spatially localized, in particular the localization of the source entails that the neutrinos are produced in \emph{entangled wave packets}, the disentanglement of the charged lepton brings in another localization scale (at which the charged lepton is  measured, absorbed or decays) which also influences the disentangled neutrino state. Wave packet localization also introduces yet another decoherence length scale $L_{coh} \propto \sigma E^2_{\nu}/\delta m^2$ where $\sigma$ is the spatial localization scale of the wave packet. For $\delta m^2 \simeq 1\,\mathrm{eV}^2$ sterile neutrinos in reactor experiments it is \emph{possible} that $L_{coh} \lesssim L_{osc}$ which would result in yet another source of decoherence and suppression of oscillations. These aspects are currently being studied and will be reported in a forthcoming study\cite{junme}.

Finally, it is worth commenting that quantum entanglement is also ubiquitous in B-meson oscillations, where
the process of ``flavor'' tagging actually disentangles the \emph{entangled} $B^0_q-\overline{B}^0_q$ state produced by $\Upsilon(4s)$ decay\cite{glashow,bert,dass}, and quantum entanglement of the $C=-1$   $\overline{B}_s\,B_s$ pair produced in the decay of the $\Upsilon(5s)$ has been invoked for a determination of the width difference\cite{soni}. Thus neutrino mixing is yet another fascinating manifestation of quantum entanglement in a system that maintains \emph{macroscopic quantum coherence} over scales of kilometers. Fascinating examples of quantum entanglement on macroscopic scales are also emerging in other unlikely systems: photosynthesis in light harvesting complexes\cite{photosynthesis} and perhaps most surprising and provocative, as a possible explanation of the ``avian compass''\cite{entanglementbirds}.

\acknowledgements The author thanks Tony Mann, Vittorio Paolone, Donna Naples and David Jasnow for enjoyable and enlightening conversations, he is indebted  to William C.  Louis III for extensive and instructive correspondence and acknowledges support from NSF through  grant award
 PHY-0852497.

\appendix

\section{\label{sec:WW} The Wigner-Weisskopf Method}\label{apsec:WW}

For completeness we give a detailed presentation of the field theoretical version of the Wigner-Weisskopf approximation as it is not widely available in the literature.

Consider a system whose Hamiltonian  $H=H_0+H_I$ where $H_0$ is the free field Hamiltonian and $H_I$ the interaction. The time evolution of states in the interaction picture
of $H_0$ is given by
\be i \frac{d}{dt}|\Psi(t)\rangle_I  = H_I(t)\,|\Psi(t)\rangle_I,  \label{intpic}\ee
where the interaction Hamiltonian in the interaction picture is
\be H_I(t) = e^{iH_0\,t} H_I e^{-iH_0\,t} \label{HIoft}\ee

This has the formal solution
\be |\Psi(t)\rangle_I = U(t,t_0) |\Psi(t_0)\rangle_I \label{sol}\ee
where   the time evolution operator in the interaction picture $U(t,t_0)$ obeys \be i \frac{d}{dt}U(t,t_0)  = H_I(t)U(t,t_0)\,. \label{Ut}\ee

Now we can expand \be |\Psi(t)\rangle_I = \sum_n C_n(t) |n\rangle \label{decom}\ee where $|n\rangle$ form a complete set of orthonormal eigenstates of $H_0$; in the quantum field theory case these are  many-particle Fock states. From eq.(\ref{intpic}) one finds the {\em exact} equation of motion for the coefficients $C_n(t)$, namely

\be \dot{C}_n(t) = -i \sum_m C_m(t) \langle n|H_I(t)|m\rangle \,. \label{eofm}\ee

\begin{figure}[ht!]
\begin{center}
\includegraphics[height=3in,width=3in,keepaspectratio=true]{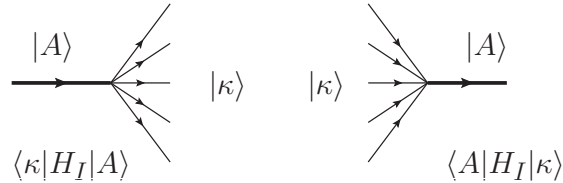}
\caption{Transitions $|A\rangle \leftrightarrow |\kappa\rangle$ in first order in $H_I$.}
\label{fig1:coupling}
\end{center}
\end{figure}

Although this equation is exact, it generates an infinite hierarchy of simultaneous equations when the Hilbert space of states spanned by $\{|n\rangle\}$ is infinite dimensional. However, this hierarchy can be truncated by considering the transition between states connected by the interaction Hamiltonian at a given order in $H_I$. Thus
consider the situation depicted in figure~\ref{fig1:coupling} where one state, $|A\rangle$, couples to a set of states $\left\{|\kappa\rangle\right\}$, which couple back   to $|A \rangle$ via $H_I$.

Under these circumstances, we have \bea \dot{C}_A(t) & = & -i \sum_{\kappa} \langle A|H_I(t)|\kappa\rangle \,C_\kappa(t)\label{CA}\\
\dot{C}_{\kappa}(t) & = & -i \, C_A(t) \langle \kappa|H_I(t) |A\rangle \label{Ckapas}\eea where the sum over $\kappa$ is over all the intermediate states coupled to $|A\rangle$ via $H_I$.

Consider the initial value problem in which at time $t=0$ the state of the system $|\Psi(t=0)\rangle = C_A(0) |A\rangle$, namely \be    C_{\kappa}(t=0) =0 .\label{initial}\ee  We can solve eq.(\ref{Ckapas}) and then use the solution in eq.(\ref{CA}) to find \bea  C_{\kappa}(t) & = &  -i \,\int_0^t \langle \kappa |H_I(t')|A\rangle \,C_A(t')\,dt' \label{Ckapasol}\\ \dot{C}_A(t) & = & - \int^t_0 \Sigma_A(t-t') \, C_A(t')\,dt' \label{intdiff} \eea where, using (\ref{HIoft}) we find \be \Sigma_A(t-t') = \sum_\kappa \langle A|H_I(t)|\kappa\rangle \langle \kappa|H_I(t')|A\rangle =  \sum_\kappa |\langle A|H_I(0)|\kappa\rangle|^2\,e^{i(E_A-E_\kappa)(t-t')} \label{sigma} \ee This integro-differential equation  with {\em memory} yields a non-perturbative solution for the time evolution of the amplitudes and probabilities. Inserting the solution for $C_A(t)$ into eq.(\ref{Ckapasol}) one obtains the time evolution of amplitudes $C_{\kappa}(t)$ from which we can compute  the time dependent probability to populate the state $|\kappa\rangle$, $|C_\kappa(t)|^2$. This is the essence of the Weisskopf-Wigner\cite{ww} non-perturbative method ubiquitous in quantum optics\cite{qoptics} and the decay formalism of $K_0-\overline{K}_0$ mixing\cite{cp}.

The hermiticity of the interaction Hamiltonian $H_I$, together with the initial conditions in eqs.(\ref{initial}) yields the unitarity condition
\be \sum_n |C_n(t)|^2 =1\,. \label{unitarity1}\ee

Equation (\ref{intdiff}) can be solved exactly via Laplace transform\cite{desiternuestro}, however, in
weak coupling, the time evolution of $C_A(t)$ determined by eq.(\ref{intdiff}) is \emph{slow} in the sense that
the time scale is determined by a weak coupling kernel $\Sigma \propto H^2_I$. This allows us to use a Markovian approximation in terms of a
consistent expansion in derivatives of $C_A$\cite{desiternuestro}. Define \be W_0(t,t') = \int^{t'}_0 \Sigma_A(t-t'')dt'' \label{Wo}\ee so that \be \Sigma_A(t-t') = \frac{d}{dt'}W_0(t,t'),\quad W_0(t,0)=0. \label{rela} \ee Integrating by parts in eq.(\ref{intdiff}) we obtain \be \int_0^t \Sigma_A(t-t')\,C_A(t')\, dt' = W_0(t,t)\,C_A(t) - \int_0^t W_0(t,t')\, \frac{d}{dt'}C_A(t') \,dt'. \label{marko1}\ee The second term on the right hand side is formally of \emph{fourth order} in $H_I$ and we see how a systematic approximation scheme can be developed. Setting \be W_1(t,t') = \int^{t'}_0 W_0(t,t'') dt'', \quad W_1(t,0) =0 \,\label{marko2} \ee and integrating by parts again, we find \be \int_0^t W_0(t,t')\, \frac{d}{dt'}C_A(t') \,dt' = W_1(t,t)\,\dot{C}_A(t) +\cdots \label{marko3} \ee leading to   \be \int_0^t \Sigma(t,t')\,C_A(t')\, dt' = W_0(t,t)\,C_A(t) - W_1(t,t)\,\dot{C}_A(t) +\cdots \label{histoint}\ee

This process can be implemented systematically resulting in higher order differential equations. Up to leading order in this Markovian approximation the equation eq.(\ref{intdiff}) becomes \be \dot{C}_A(t) \left[1- W_1(t,t)\right] + W_0(t,t) C_A(t) =0 \label{markovian}\ee with the result \be C_A(t) = e^{-i\int_0^t \mathcal{E}(t')dt'},\quad \mathcal{E}(t) = \frac{-i\,W_0(t,t)}{1-W_1(t,t)} \simeq -i\,W_0(t,t)\left[1+W_1(t,t)+\cdots\right] \label{solumarkov}\ee To leading order in the interaction ($\mathcal{O}(H_I^2)$) we keep  $\mathcal{E}(t) = -i\,W_0(t,t) $.  Note that in general $\mathcal{E}(t)$ is complex. In the long time limit and using the representation (\ref{sigma}) we find
\be \int^{\infty}_0 \Sigma_A(\tau)d\tau = i\, \sum_\kappa \frac{|\langle A|H_I(0)|\kappa\rangle|^2}{(E_A-E_\kappa + i0^+)} \equiv i\,\Delta E_A + \frac{\Gamma_A}{2} \label{sigmaasi}\ee where
\be \Delta E_A = \mathcal{P} \sum_\kappa \frac{|\langle A|H_I(0)|\kappa\rangle|^2}{(E_A-E_\kappa )} \label{deltaEA}\ee is the energy shift in agreement with second order perturbation theory, and
\be \Gamma_A = 2\pi\, \sum_\kappa |\langle A|H_I(0)|\kappa\rangle|^2 \, \delta(E_A-E_\kappa) \label{widthA} \ee this result for the width is in agreement with Fermi's Golden rule. Finally, in the Markovian approximation the Wigner-Weisskopf method yields
\be  C_A(t) =  C_A(0)\, e^{-i\Delta E_A \,t}\,e^{-\frac{\Gamma_A}{2}\,t}\,. \label{CAoft}\ee This solution agrees with the \emph{exact} solution via Laplace transform\cite{desiternuestro}\footnote{Here we neglect wave function renormalization as it is not relevant for the discussion.}. Inserting this result into equation (\ref{Ckapasol}) we find
\bea  C_{\kappa}(t)   = &&  -i\,  C_A(0)\,\langle\kappa |H_I(0)|A\rangle \,\int_0^t  e^{-i(E^r_A-E_\kappa-i\frac{\Gamma_A}{2})t'}\,dt' \nonumber \\ = &&  -   C_A(0)\,\langle \kappa |H_I(0)|A\rangle \,\Bigg[\frac{1-e^{-i(E^r_A-E_\kappa-i\frac{\Gamma_A}{2})\,t}}{E^r_A-E_\kappa-i\frac{\Gamma_A}{2}} \Bigg] \label{Ckapasol2}\eea where $E^r_A = E_A+\Delta E_A$ is the renormalized energy. The Schroedinger picture state $|\Psi(t)\rangle_S = e^{-iH_0t}|\Psi(t)\rangle_I$ is finally given by
\be |\Psi(t)\rangle_S = C_A(0)\Bigg\{ e^{-iE^r_A\,t}\,e^{-\frac{\Gamma_A}{2}\,t}|A\rangle - \sum_\kappa \langle \kappa |H_I(0)|A\rangle \,\Bigg[\frac{1-e^{-i(E^r_A-E_\kappa-i\frac{\Gamma_A}{2})\,t}}{E^r_A-E_\kappa-i\frac{\Gamma_A}{2}} \Bigg]\,e^{-iE_\kappa\,t}\,|\kappa\rangle \Bigg\} \,.\label{schpicstate}\ee For $t\gg \tau_A=1/\Gamma_A$ the asymptotic state becomes
\be \big|\Psi(t\gg \tau_A)\big\rangle_S =   -  C_A(0)\sum_\kappa \frac{\langle \kappa |H_I(0)|A\rangle\,e^{-iE_\kappa\,t}} {\Big[E^r_A-E_\kappa-i\frac{\Gamma_A}{2}\Big]} ~|\kappa\rangle   \,.\label{schpicstateasy}\ee

\end{document}